\documentclass[useAMS,usenatbib,fleqn]{mn2e}
\usepackage{graphicx}
\usepackage{amsmath,amssymb}
\usepackage{bm}
\usepackage{natbib,aas_macros}
\title[Orbital eccentricity distribution of the halo stars]
{A new test for the Galactic formation and evolution 
-- prediction for the orbital eccentricity distribution of the halo stars}
\author[K. Hattori and Y. Yoshii]{K. Hattori$^{}$\thanks{E-mail:
khattori@ioa.s.u-tokyo.ac.jp (KH); yoshii@ioa.s.u-tokyo.ac.jp (YY)} and Y. Yoshii$^{}$\footnotemark[1]\\
$^{}$Institute of Astronomy, School of Science, University of Tokyo, Tokyo 181-0015, Japan}

\begin{document}

\date{Accepted 2010 May 11. Received 2010 May 10; in original form 2009 December 10}

\pagerange{\pageref{firstpage}--\pageref{lastpage}} \pubyear{20XX}

\maketitle

\label{firstpage}

\begin{abstract}
We present theoretical calculations for the differential distribution of stellar 
orbital eccentricity in a galaxy halo, assuming that the stars constitute a spherical, 
collisionless system in dynamical equilibrium with a dark matter halo.  In order to 
define the eccentricity $e$ of a halo star for given energy $E$ and angular momentum 
$L$, we adopt two types of gravitational potential, such as an isochrone potential 
and a Navarro-Frenk-White potential, that could form two ends covering in-between any 
realistic potential of dark matter halo. 
Based on a distribution function of the form $f(E,L)$ that allows  
constant anisotropy in velocity dispersions characterized by a parameter $\beta$, 
we find that the eccentricity distribution is a monotonically increasing function 
of $e$ for the case of highly radially anisotropic velocity dispersions 
($\beta \gtrsim 0.6$), while showing a hump-like shape 
for the cases from radial through tangential velocity anisotropy ($\beta \lesssim 0.6$). 
We also find that when the velocity anisotropy agrees with that observed 
for the Milky Way halo stars ($\beta \simeq 0.5-0.7$),
a nearly linear eccentricity distribution of $N(e) \propto e$ results 
at $e \lesssim 0.7$, largely independent of the potential adopted. 
Our theoretical eccentricity distribution would 
be a vital tool of examining 
how far out in the halo the dynamical equilibrium 
has been achieved, through comparison with kinematics of halo stars sampled at greater 
distances. Given that large surveys of the SEGUE and Gaia projects would be in progress, 
we discuss how our results would serve as a new guide in exploring the formation and 
evolution of the Milky Way halo. 
\end{abstract}


\section{Introduction} \label{intro}

Studies of large-scale structures in the universe and fluctuations in the 
cosmic microwave background strongly favor a $\Lambda$-cold dark matter 
($\Lambda$CDM) cosmology (e.g., Cole et al. 2005; Dunkley et al. 2009). The formation 
of structures in this cosmology is a process of hierarchical clustering, in 
the sense that numerous CDM lumps cluster gravitationally and merge together 
to form larger structures (White \& Rees 1978; Blumenthal et al. 1984).  

Dark halos of galaxy systems are similarly formed via clustering of 
subhalos as a result of CDM agglomerations that reach the maximum expansion 
then turn around to collapse in the background expanding medium, but a 
detailed process leading to the halo formation from primordial density 
fluctuations is highly nonlinear and is not as simple as the formation of 
larger structures in the universe (e.g., for review see Ostriker 1993 and 
Bertschinger 1998). 

High-resolution $\Lambda$CDM simulations for the halo formation generically 
show that mergers and collisions of subhalos induce the overall collapse and 
virialize the inner region of host halo, while surviving subhalos orbit as 
separate entities within the inner virialized region of halo (e.g., Moore 
et al. 1999; Ghigna et al. 2000; Helmi, White \& Springel 2003; Valluri et al. 2007). 
A majority of stars formed through this build-up of halo are expected to 
have also experienced the redistribution of energy and momentum that drives 
the phase mixing or violent relaxation towards the dynamical equilibrium 
(Lynden-Bell 1967). This leads to an idea that a stellar halo, which can be 
regarded as a collisionless system, holds the dynamical information just 
after the last violent relaxation in forming the halo. 

We then take an approach to find out the relics of the formation of the Milky Way 
halo from the kinematics of halo stars.  Among many of their kinematic properties 
available at present and in the near future, the differential distribution 
of stellar orbital eccentricity $N(e)$ seems to be of special importance. 
The orbital eccentricity of a star is a quasi-adiabatic invariant (Eggen, 
Lynden-Bell \& Sandage 1962; Lynden-Bell 1963) and is unaffected by the small 
and slow variation of the gravitational potential that might have occurred 
after the major formation of halo stars.  It is therefore most likely that 
the shape of $N(e)$ has been conserved until present.  With this consideration, 
comparing the observed shape of $N(e)$ for halo stars with the theoretical one 
for the halo in dynamical equilibrium, we could explore 
how far out 
in the halo the dynamical equilibrium was achieved. Consequently, $N(e)$ 
serves as a new test of halo formation scenario in a $\Lambda$CDM cosmology.

As a useful way to derive $N(e)$ theoretically, 
we consider the orbit of halo stars in assumed gravitational potentials of the halo. 
In section 2, we 
present our formulation to calculate $N(e)$ under some plausible assumptions 
for the halo, and apply it to two extreme gravitational potentials of academic 
interest. The results for realistic cases are shown for the isochrone potential 
and for the Navarro-Frenk-White (NFW) potential in section 3.  We summarize 
the results and discuss the prospects of investigating the formation and evolution 
of the Milky Way halo in section 4.

\section[]{Formulation} \label{method}
We assume that the halo stars constitute a spherical, collisionless system 
in dynamical equilibrium with a dark halo.  Since the dark matter is known 
to dominate the total mass of the galaxy system, the motion of halo stars is 
governed by the gravitational potential of dark halo.

\subsection{Stellar orbital eccentricity in a model halo} \label{e}

When a spherical halo potential $V(r)$ is given with respect to the galaxy 
center, the energy $E$ and the angular momentum $L$ of a star at the position 
${\bm r}$ with the velocity ${\bm v}$ are written respectively as 
\begin{equation}
E=\frac{1}{2} {\bm v}^2 + V(r), \hspace{2ex} {\rm and} \hspace{2ex}
L = |{\bm L}| = \left| {\bm r}\times {\bm v} \right| ,  
\end{equation}
where $r=|{\bm r}|$. The orbital eccentricity of a star is practically 
defined as 
\begin{equation} \label{e-def}
e \equiv \frac{ r_{\rm apo} - r_{\rm peri} }{ r_{\rm apo} + r_{\rm peri} } ,
\end{equation}
where $r_{\rm apo}$ and $r_{\rm peri}$ are the apo- and peri-centric distances, 
respectively, and are given by two real solutions ($r_{\rm apo}>r_{\rm peri}$) 
of the following equation:
\begin{equation} \label{eq_r1r2}
E = V(r) + \frac{L^2}{2 r^2} \equiv V_{\rm{eff}} (L;r) . 
\end{equation}
It is evident from equations (\ref{e-def}) and (\ref{eq_r1r2}) that a pair of 
$(E,L)$ has a one-to-one correspondence to $(E,e)$, but there is a region 
of $(E,L)$ in which two real solutions are not allowed and thus, except 
for the case of circular orbits, the eccentricity cannot be defined.  
Since such unbound orbits do not form a steady population of stellar halo, 
we neglect them and exclusively consider stars with bound orbits.  Constraints 
on $E$ and $L$ that allow bound orbits are presented in Appendix A.

\subsection{Differential distribution of stellar orbital eccentricity} \label{d}

Let $f({\bm r},{\bm v})$ be the distribution function of halo stars, then the  
number of halo stars in a phase space volume $d^3 {\bm r} d^3 {\bm v}$ centered 
at $({\bm r}, {\bm v})$ is given by $f({\bm r}, {\bm v}) d^3{\bm r}d^3{\bm v}$. 
According to the 
strong
Jeans theorem, the distribution function should be expressed 
in terms of isolating integrals only (Lynden-Bell 1960, 1962).  For a spherical 
system that is invariant under rotation, it takes a form of either $f(E)$ or 
$f(E,L)$, depending on whether the stellar velocity dispersion is isotropic or 
anisotropic, respectively.

The velocity dispersion observed for halo stars is radially anisotropic (e.g., 
Yoshii \& Saio 1979; Gilmore, Wyse, \& Kuijken 1989).  Furthermore, recent 
observations for halo stars within the distance of $10 \hspace{1ex} {\rm kpc}$ 
away from us show that the shape of velocity ellipsoid is constant and its 
principal axes are well aligned with the spherical coordinates (Carollo et al. 
2007; Bond et al. 2009). If we extrapolate this fact to a whole system, one  
simple form of the distribution function is 
\begin{equation}\label{f_beta}
    f(E,L) = 
    \begin{cases}
        g(E) L^{-2 \beta} , & \text{if $(E,L)$ is `allowed'} \\
        0 ,                  & \text{\rm otherwise,}  
    \end{cases}
\end{equation}
where $g(E)$ is a function of $E$ (e.g., Binney \& Tremine 2008).  
Here, $\beta$ is a constant value of 
velocity anisotropy parameter defined as 
\begin{equation}
\beta \equiv 1 - \frac{\sigma_{\rm t}^2}{\sigma_r^2}, 
\end{equation}
where $\sigma_{r}$ is the radial velocity dispersion and $\sigma_{t}$ is the 
tangential velocity dispersion projected onto the spherical $\theta -\phi$ 
surface.  
Although $\beta$ is about $0.5-0.7$ observationally 
(e.g., Bond et al. 2009; Smith et al. 2009; Carollo et al. 2010), 
we will use it as a constant parameter below. 

By changing the variables and integrating over the spherical coordinates, 
the number of stars in $d^3 {\bm r} d^3 {\bm v}$ reduces to
\begin{equation} \label{with_L}
N(E,L)dEdL^2 = 4\pi^2 g(E) L^{-2\beta} T_r dE dL^2 , 
\end{equation}
with the radial period of stellar orbit given by 
\begin{equation}\label{Tr}
T_r (E,L) \equiv \oint \frac{dr}{v_r} 
= 2 \int_{r_{\rm peri}}^{r_{\rm{apo}}} \frac{dr}{\sqrt{2 \left[ E-V_{\rm{eff}}(L;r) \right]}}. 
\end{equation}
Since $L^2$ is a function of $E$ and $e$, we here introduce the $E$-dependent 
differential eccentricity distribution as
\begin{equation} \label{general-n}
n_{\beta}(E,e) = 4 \pi^2 L^{-2 \beta} \cdot T_r 
\cdot \left|\left(\frac{\partial L^2}{\partial e}\right)_{E}\right| .
\end{equation}
We then express the differential eccentricity distribution as
\begin{equation} \label{general-N}
N_{\beta}(e) = \int_{{\rm allowed} \; E} g(E) n_{\beta}(E,e) dE . 
\end{equation}
It is apparent from this equation that $N_{\beta}(e)$ is a weighted sum of 
$n_{\beta}(E,e)$ with a weight function of $g(E)$.  Thus, once the gravitational 
potential $V(r)$ and the velocity anisotropy parameter $\beta$ are specified, 
we can formally obtain $n_{\beta}(E,e)$, and also $N_{\beta}(e)$ after integrating 
$n_{\beta}(E,e)$ over $E$ with its appropriate weight.

\subsection{Extreme cases of mass distribution} \label{extreme}

In this subsection, mostly for pedagogical purpose, we consider two extreme 
cases of mass distribution such as the point mass at the center and the 
homogeneous distribution in the truncated sphere.  These cases allow analytic 
expression of $n_{\beta}(E,e)$, and because it is separable in $E$ and $e$, 
$N_{\beta}(e)$ can also be obtained except for its normalization.  Therefore, 
these cases are helpful to understand the results for any more realistic cases.

\subsubsection{Central point mass} \label{central}
\begin{figure*}
\begin{center}
	\includegraphics[width=\columnwidth]{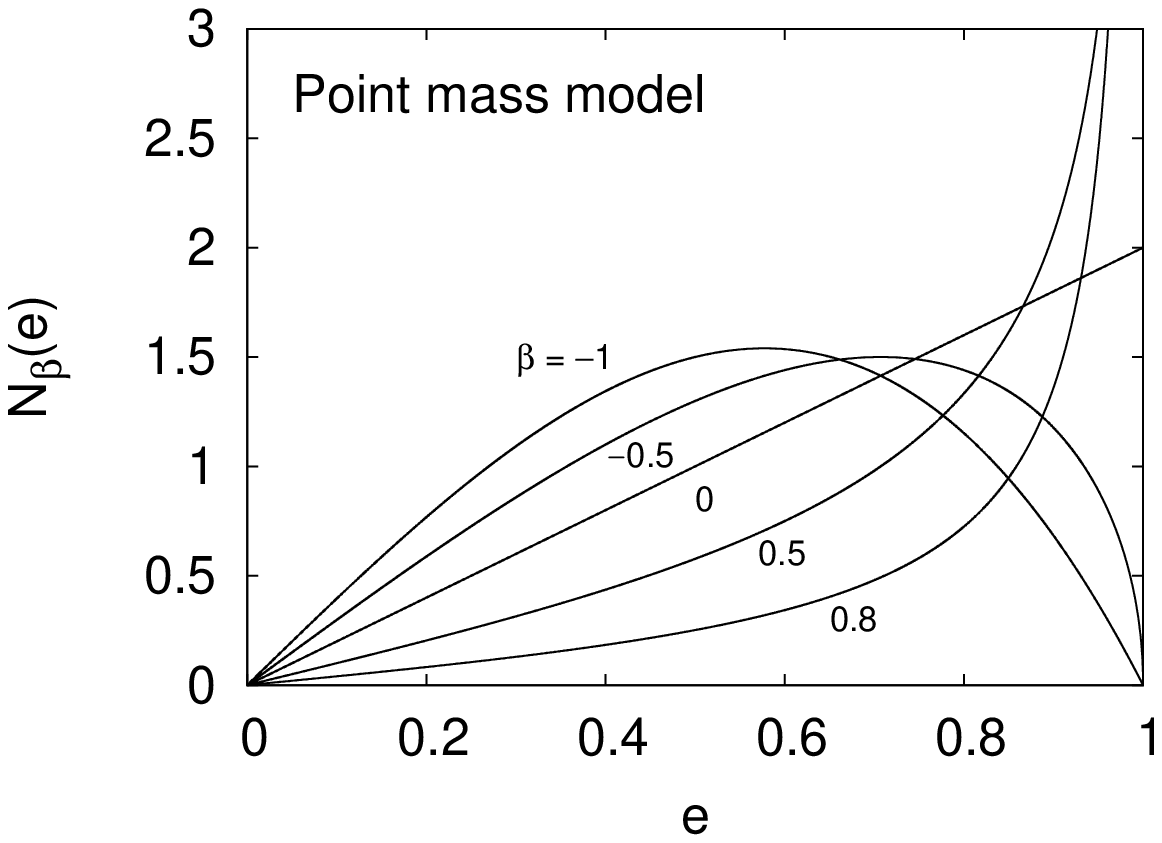}
	\includegraphics[width=\columnwidth]{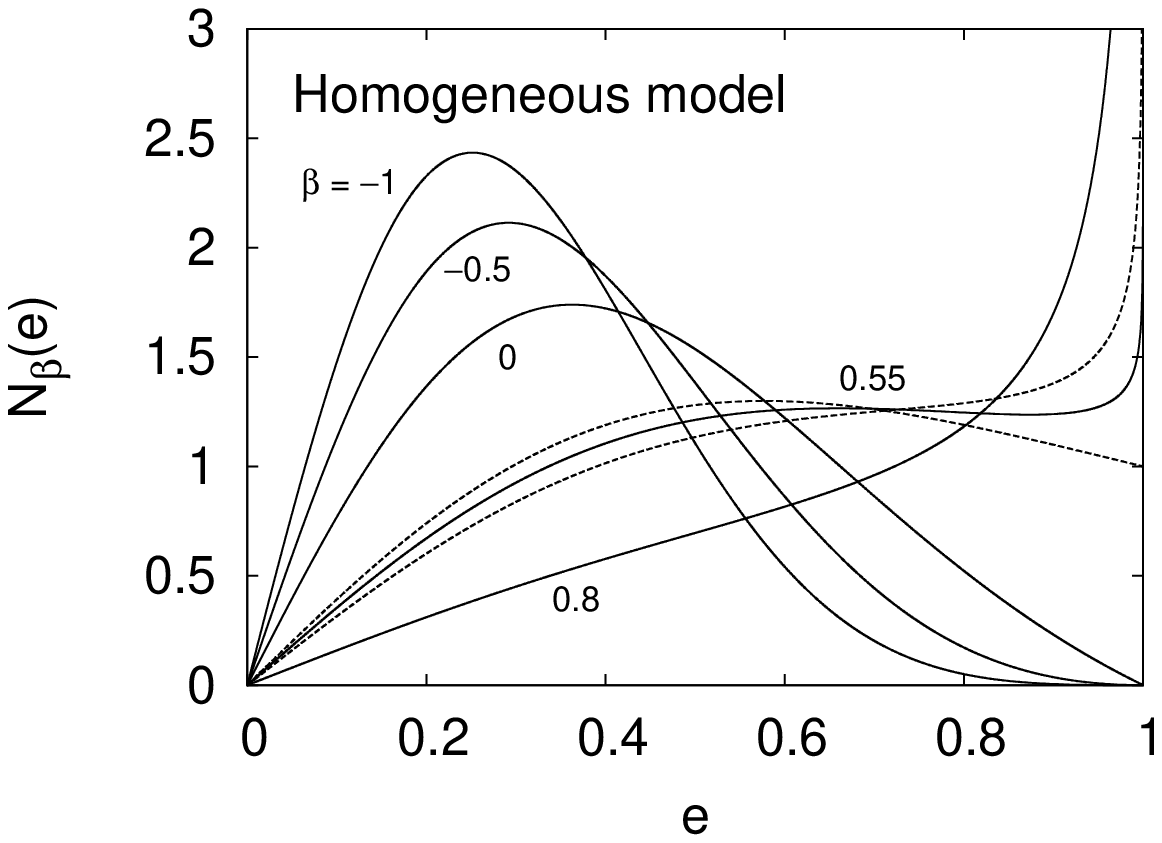}\\
	\caption{Differential distribution of stellar orbital eccentricity $N_{\beta}(e)$
in two extreme cases of mass distribution, such as the point mass model on the left panel  
and the homogeneous model on the right panel.  The results are shown by 
lines for several values of velocity anisotropy parameter $\beta$.  If $N_{\beta}(e)$ 
near $e=1$ sensitively changes at some particular value of $\beta$, 
the results for $\beta\pm 0.05$ are additionally shown by dotted lines for the 
purpose of illustrating its sensitivity.  
Note that $N_{\beta}(e)$ is normalized such that $\int_{0}^{1} N_{\beta}(e)de=1$. } 
\label{Fig Kepler+IHO-beta}
\end{center}
\end{figure*}

The gravitational potential arising from the central point mass is Keplerian 
and is given by 
\begin{equation}
V(r) =  - \frac{GM}{r} , 
\end{equation}
where $M$ is the total mass of dark halo and $G$ is the gravitational constant. 
For bound orbits with $E < 0$, there are two real and positive solutions for 
equation (\ref{eq_r1r2}), or equivalently,
\begin{equation}
(-2E)r^2 - 2GMr + L^2 = 0 .  
\end{equation}
The orbital eccentricity is expressed in terms of $(E,L)$ as 
\begin{equation}\label{1-e}
e =  \sqrt{1 - \frac{(-2E)L^2}{{(GM)}^2}} , 
\end{equation}
and the other relevant quantities are neatly expressed as 
\begin{equation}
T_r = \frac{2 \pi GM}{ {(-2E)}^{\frac{3}{2}} }, \hspace{2ex} 
{\rm and} \hspace{2ex} L^2 = \frac{{(GM)}^2}{-2E} (1-e^2). 
\end{equation}
Substitution of these quantities in equations (\ref{general-n}) 
and (\ref{general-N}) gives the $E$-dependent 
differential eccentricity distribution
\begin{equation}
n_{\beta} (E,e) 
= 16 \pi^3 {(GM)}^{3-2\beta} {(-2 E)}^{\beta - \frac{5}{2} } \frac{e}{{(1-e^2)}^{\beta}} ,
\end{equation}
and the differential eccentricity distribution
\begin{equation}
N_{\beta} (e) 
= 16 \pi^3 {(GM)}^{3-2\beta} \left[\int g(E){(-2 E)}^{\beta - \frac{5}{2} }dE\right] 
\frac{e}{{(1-e^2)}^{\beta}} .
\end{equation}
Since $N_{\beta}(e)\propto n_{\beta} (E,e)$, we normalize $N_{\beta}(e)$ 
such that 
$\int_{0}^{1} N_{\beta}(e)de=1$, and write    
\begin{equation}
{\rm normalized}\;\; N_{\beta}(e)=2(1-\beta) \frac{e}{{(1-e^2)}^{\beta}}, \hspace{2ex} (\beta \neq 1) .
\end{equation} 
The results of $N_{\beta}(e)$ for several values of $\beta$ are shown 
on the left panel of Figure \ref{Fig Kepler+IHO-beta}.  
For the case of $\beta=0$ (isotropic velocity dispersion), $N_{\beta}(e)$ is 
exactly proportional to $e$ (Binney \& Tremaine 2008) and 
we call it the linear eccentricity distribution.  For $0<\beta<1$ (radially anisotropic 
velocity dispersion), $N_{\beta}(e)$ is a rapidly increasing function of $e$ 
with a peak always at $e=1$.  On the other hand, for $\beta<0$ (tangentially 
anisotropic velocity dispersion), $N_{\beta}(e)$ shows a hump-like $e$-distribution 
around a single peak at $e=(1-2\beta)^{-1/2}$.  

We should notice that the linear trend of $N_{\beta}\propto e$ prevails in a 
range of $0<e<0.3$ regardless of $\beta$, while the behavior of 
$N_{\beta}(e)$ is very sensitive to $\beta$ in a range of $0.6<e<1$ and 
the difference there clearly shows up.

\subsubsection{Truncated homogeneous sphere} \label{homo}
 
A homogeneous density distribution within truncated sphere is expressed as    
\begin{equation}
    \rho(r) = 
    \begin{cases}
        \frac{3M}{4 \pi r_{\rm t}^3} , & \text{if $r < r_{\rm t}$} \\
        0 ,                      & \text{otherwise,} 
    \end{cases}
\end{equation}
where $M$ is the total mass of dark halo and $r_{\rm t}$ is the truncation radius. 
The gravitational potential arising from this density distribution is given by  
\begin{equation}
     V(r) =
     \begin{cases}
     - \frac{3GM}{2 r_{\rm t}} + \frac{GM}{2 r_{\rm t}} {\left( \frac{r}{r_{\rm t}} \right) }^2 , 
     & \text{if $r < r_{\rm t}$} \\ 
     - \frac{GM}{r} , & \text{otherwise.}
     \end{cases}
\end{equation}
We consider only stars with $E<E_{\rm t}\equiv-GM/r_{\rm t}$, 
which guarantees the stars to be confined inside the truncated radius $r_{\rm t}$.  
Thus, bound orbits 
within the truncated sphere are allowed if $E_{\rm min}<E<E_{\rm t}$ where we note  
$E_{\rm min} \equiv (3/2)E_{\rm t}$. In this limited range of $E$, there are two 
real and positive solutions for equation (\ref{eq_r1r2}), or equivalently,
\begin{equation}
GMr_{\rm t} {\left( \frac{r}{r_{\rm t}} \right)}^4 
- 2 \left( E - E_{\rm min} \right) r_{\rm t}^2 {\left( \frac{r}{r_{\rm t}} \right)}^2+ L^2 = 0 , 
\end{equation}
if and only if 
\begin{equation}
0 < D < 1, 
\end{equation}
where 
\begin{equation}
D = \frac{GML^2}{r_{\rm t}^3  {\left( E -E_{\rm min} \right)}^2 } . 
\end{equation}
The orbital eccentricity is expressed in terms of $D$ as 
\begin{equation}\label{2-Ie}
e = \sqrt{\frac{1-\sqrt{D}}{1+\sqrt{D}}} , 
\end{equation}
and the other relevant quantities are expressed in terms of $(E,L)$ as 
\begin{equation}
T_r = \pi \sqrt{ \frac{r_{\rm t}^3}{GM} }, \hspace{2ex} 
{\rm and} \hspace{2ex}
 L^2 = \frac{r_{\rm t}^3}{GM} {\left( E - E_{\rm min} \right)}^2 { \left(\frac{1-e^2}{1+e^2}\right) }^2 .
\end{equation}
Consequently, we obtain
\begin{equation} \label{nIHO}
n_{\beta}(E,e) = 32 \pi^3 {\left( \frac{r_{\rm t}^3}{GM} \right)}^{\frac{3}{2} - \beta} 
{\left( E - E_{\rm min} \right)}^{2-2\beta} \frac{e {(1-e^2)}^{1-2\beta}}{{(1+e^2)}^{3-2\beta}} , 
\end{equation}
and 
\begin{multline}
N_{\beta}(e) = 32 \pi^3 {\left( \frac{r_{\rm t}^3}{GM} \right)}^{\frac{3}{2} - \beta} 
\left[ \int g(E) {\left( E - E_{\rm min} \right)}^{2-2\beta}dE \right] \\
\times
\frac{e {(1-e^2)}^{1-2\beta}}{{(1+e^2)}^{3-2\beta}} . 
\end{multline}
As in the point mass model, since $N_{\beta}(e)\propto n_{\beta} (E,e)$, 
we normalize $N_{\beta}(e)$ such that $\int_{0}^{1} N_{\beta}(e)de=1$, and write  
\begin{equation}
{\rm normalized}\;\; N_{\beta}(e)=8(1-\beta) 
\frac{e {(1-e^2)}^{1-2\beta}}{{(1+e^2)}^{3-2\beta}} \hspace{2ex}, (\beta \neq 1) .
\end{equation}

The results of $N_{\beta}(e)$ for several values of $\beta$ are shown 
on the right panel of Figure \ref{Fig Kepler+IHO-beta}.  
For $\beta<0.5$, $N_{\beta}(e)$ shows a hump-like $e$-distribution with a single 
peak at 
\begin{equation}
e_{\rm peak}= \sqrt{\frac{ 4(1-\beta) - \sqrt{13 - 32\beta + 16\beta^2} }{3} } . 
\end{equation}
For $0.5<\beta<1-\sqrt{3}/4$, however, $N_{\beta}(e)$ has two local maxima such as 
a broad peak at $e=e_{\rm peak}$ and a sharp peak at $e=1$.  Overall behavior 
monotonically increases with $e$ in a range of $0<e<e_{\rm peak}$, and is kept more 
or less flat in the range of $e_{\rm peak}<e<1$.  
For $1-\sqrt{3}/4<\beta<1$, $N_{\beta}(e)$ 
is a rapidly increasing function of $e$. 

For a given value of $\beta$, 
$N_{\beta}(e)$ is more weighted at smaller $e$ in the homogeneous model, 
when compared with the point mass model.  In particular, for $\beta=0$ 
(isotropic velocity dispersion), $N_{\beta}(e)$ shows a broad hump-like $e$-distribution 
around a peak at $e_{\rm peak}=0.36$ in the homogeneous model, while showing an exactly 
linear $e$-distribution in the point mass model.  This sensitivity, though between 
two extreme cases, could be used to discriminate the likely mass distribution in 
more realistic cases to be considered in section 3.

\subsection{Effect of central mass concentration} \label{effect}

In the cases of central point mass and truncated homogeneous sphere, the shape of 
$n_{\beta}(E,e)$ is the same as $N_{\beta}(e)$, because $n_{\beta}(E,e)$ is separable 
in $E$ and $e$ and thus the shape of $N_{\beta}(e)$ is unaffected by $g(E)$ in 
equation (\ref{f_beta}).  This property generally holds when the density 
distribution in the truncated sphere is given by $\rho (r) \propto 1/r^{\gamma}$  
(see Appendix B).  The homogeneous model in section \ref{homo} corresponds to 
$\gamma = 0$.

\begin{figure}
\begin{center}
	\includegraphics[width=\columnwidth]{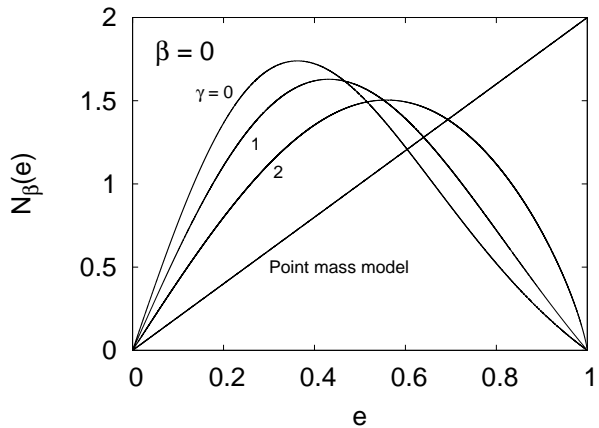}
        \caption{Differential distribution of stellar orbital eccentricity $N_{\beta}(e)$ 
for four types of model potentials, when the stellar velocity dispersion is 
isotropic ($\beta=0$).  Considered are the homogeneous model 
($\gamma =0$, section \ref{homo}), the linear potential model ($\gamma =1$, Appendix B.1), 
the singular isothermal model ($\gamma =2$, Appendix B.2), and the point mass model 
(section \ref{central}), in order of increasing the central mass concentration.  
There is a clear trend that the $e$-distribution is peaked at larger $e$ as the halo mass 
is more centrally concentrated. Note that the distribution is normalized such that 
$\int_{0}^{1} N_{\beta}(e)de=1$. }
\label{Fig various-models}
\end{center}
\end{figure}

Using the cases of $\gamma =1$ (linear potential model) and $\gamma =2$ (singular 
isothermal model) that are intermediate between two extreme cases considered above, 
we can examine how $N_{\beta}(e)$ depends on the central mass concentration.  
As shown in Figure \ref{Fig various-models} for $\beta = 0$, there is a clear 
trend that the $e$-distribution is peaked at larger $e$ as the halo mass is more 
centrally concentrated.  This trend is also true regardless of the value of $\beta$ 
and is helpful in interpreting the results of more realistic models in the next section.

\section{Eccentricity distribution of halo stars} \label{example}

Our formulation in the previous section can apply to more general cases of mass 
distribution, including the isochrone model and the NFW model that could form two 
ends covering in-between any realistic cases of mass distribution of dark halo.

\subsection{Energy-dependent eccentricity distribution $n_{\beta}(\varepsilon,e)$ for the isochrone model} \label{iso}

The gravitational potential of the isochrone model (H\'enon 1959) is given by 
\begin{equation}
V(r) = - \frac{GM}{b + \sqrt{b^2 + r^2}},
\end{equation}
where $M$ is the total mass and $b$ is the scale length parameter.  Obviously, 
the asymptotic form in the limit of $r\gg b$ or $r\ll b$ approaches 
the point mass model or the homogeneous model, respectively. 
Thus, this model, though not explaining the flat rotation curve of the galaxy 
disk at greater distances from the galaxy center, is important to study the 
intermediate case of mass distribution by adjusting the scale size of the central 
core.  Furthermore, the isochrone model is particularly valuable, because fully 
analytic expression of $n_{\beta}(E,e)$ can be obtained.

Provided $b \neq 0$, we define useful dimensionless variables and effective 
potential as follows: 
\begin{equation}\label{var_iso}
x \equiv \frac{r}{b}, 
\hspace{2ex} \varepsilon \equiv \frac{2bE}{GM}, 
\hspace{2ex} \lambda \equiv \frac{2 L^2}{bGM}, 
\end{equation}
and 
\begin{equation}
\Phi_{\rm eff}(\lambda;x) \equiv 
\frac{2b}{GM} \left[ V(r) + \frac{L^2}{2 r^2} \right] 
= -\frac{2}{1+\sqrt{1+x^2}} + \frac{\lambda}{2 x^2} .
\end{equation}
Equation (\ref{eq_r1r2}) then reads  
\begin{equation} \label{eqx}
\varepsilon x^2 + 2\sqrt{1+x^2} - \left( 2+ \frac{\lambda}{2} \right) = 0. 
\end{equation}
This equation has two real and positive solutions if and only if 
\begin{equation}
-1 < \varepsilon < 0 \hspace{2ex} {\rm and} \hspace{2ex}
 0 < \lambda < \lambda_{\rm cir} \equiv \frac{2{(1+\varepsilon)}^2}{-\varepsilon} .
\end{equation}
We denote the two solutions $x_{\rm peri}$ and $x_{\rm apo}$, and use of them 
gives the relevant quantities in terms of $\varepsilon$ and $e$: 
\begin{equation}
 T_r = 2 \pi \sqrt{\frac{b^3}{GM}} (-\varepsilon)^{-\frac{3}{2}},  
\end{equation}
\begin{multline}
 L^2 = \frac{bGM}{2} 
\left( -4 - \frac{2}{\varepsilon} 
\right. \\
\left. 
+ \frac{1}{\varepsilon e^2}
\left[ (1+e^4) - (1+e^2) \sqrt{{(1-e^2)}^2 + 4 \varepsilon^2 e^2} \right] \right) , 
\end{multline}
and 
\begin{multline}
 {\left( \frac{\partial L^2}{\partial e} \right)}_{E} 
= - bGM {(-\varepsilon)}^{-1} \\
\times
\frac{(1-e^2)}{e^3}
\left[ \frac{1 + 2 \varepsilon^2 e^2 + e^4}
{\sqrt{ {(1-e^2)}^2 + 4 \varepsilon^2 e^2 }} - (1 + e^2) \right]. 
\end{multline} 
Consequently, after tedious algebra, we succeed for the first time 
to obtain analytic expression of $n_{\beta}(\varepsilon,e)$ as follows:  
\begin{multline} \label{Niso_nonzero}
n_{\beta}(\varepsilon,e) = 
8 \pi^3 \sqrt{b^5 G M} 
{(-\varepsilon)}^{-\frac{5}{2}} \\
\times
\frac{(1-e^2)}{e^3}
\left[ 
\frac{1 + 2 \varepsilon^2 e^2 + e^4}{ \sqrt{ {(1-e^2)}^2 + 4 \varepsilon^2 e^2 } }
 - (1 + e^2) 
\right] 
\\
\times 
{\left[ 
\frac{bGM}{2} 
\left( -4 - \frac{2}{\varepsilon} 
\right. \right.} 
\\
{\left. \left.
+ \frac{1}{\varepsilon e^2}
\left[ 
(1+e^4) - (1+e^2) \sqrt{ {(1-e^2)}^2 + 4 \varepsilon^2 e^2 } 
\right]
\right) 
\right]}^{-\beta} . 
\end{multline}
We see that $n_{\beta}(\varepsilon,e)$ is not separable in $\varepsilon$ 
and $e$.  Therefore, unlike the point mass and homogeneous models, the shape of 
$n_{\beta}(\varepsilon,e)$ depends on $\varepsilon$ as well as $\beta$.  
Accordingly, derivation of $N_{\beta}(e)$ needs full numerical integration 
of $n_{\beta}(\varepsilon,e)$ over $\varepsilon$ with the weight function 
$g(\varepsilon)$ specified.

When $\beta=0$, by taking a limit of $\varepsilon$, we obtain  
\begin{equation}\label{varepsilon0}
\lim_{\varepsilon \to -0} n_{\beta=0}(\varepsilon,e) 
\propto e {(-\varepsilon)}^{-\frac{5}{2}} ,
\end{equation}
and
\begin{equation}\label{varepsilon-1}
\lim_{\varepsilon \to -1} n_{\beta=0}(\varepsilon,e) \propto  
\frac{e(1-e^2)}{{(1+e^2)}^3} (1 + \varepsilon). 
\end{equation} 
These shapes of $e$-distribution exactly coincide with those in equations 
(\ref{varepsilon0}) and (\ref{varepsilon-1}), respectively. As understood 
from the definition of $\varepsilon \;[\equiv 2bE/(GM)]$, the limit of 
$\varepsilon \to 0$ corresponds to $b \to 0$ with $E$ and $M$ fixed, 
which is equivalent to taking a limit to the point mass model.  
Likewise, the limit of $\varepsilon \to -1$ corresponds to $b \to \infty$, 
otherwise such limit of $\varepsilon$ is not attained with $E$ and $M$ 
fixed, which is equivalent to taking a limit to the homogeneous model. 

\begin{figure*}
\begin{center}
	\includegraphics[width=\columnwidth]{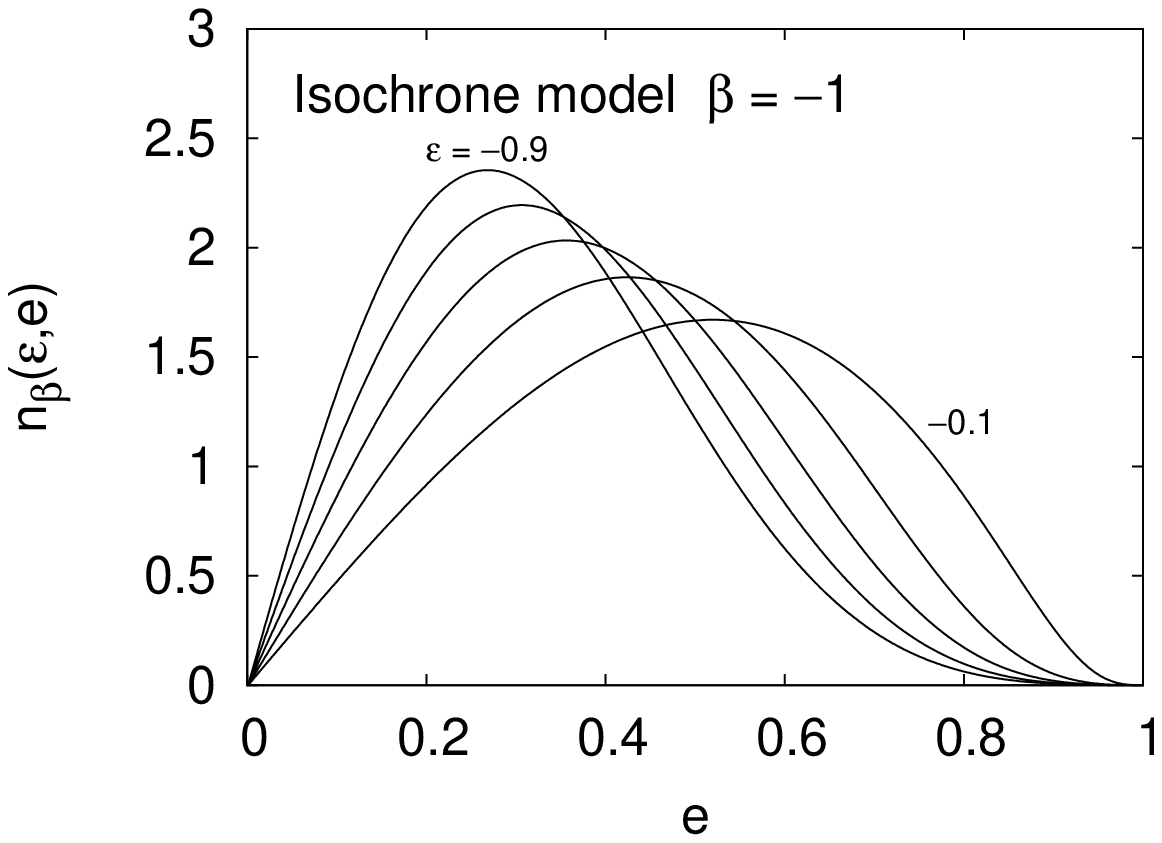}
	\includegraphics[width=\columnwidth]{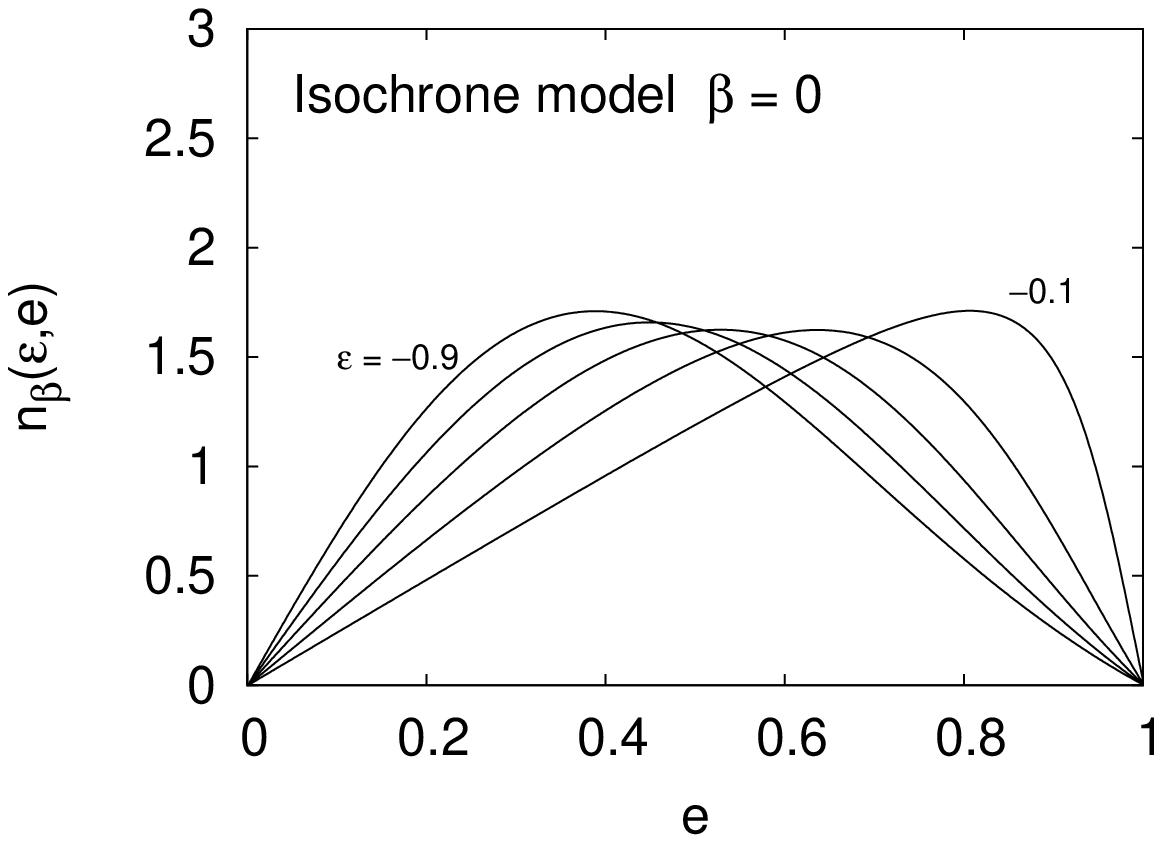}\\
	\includegraphics[width=\columnwidth]{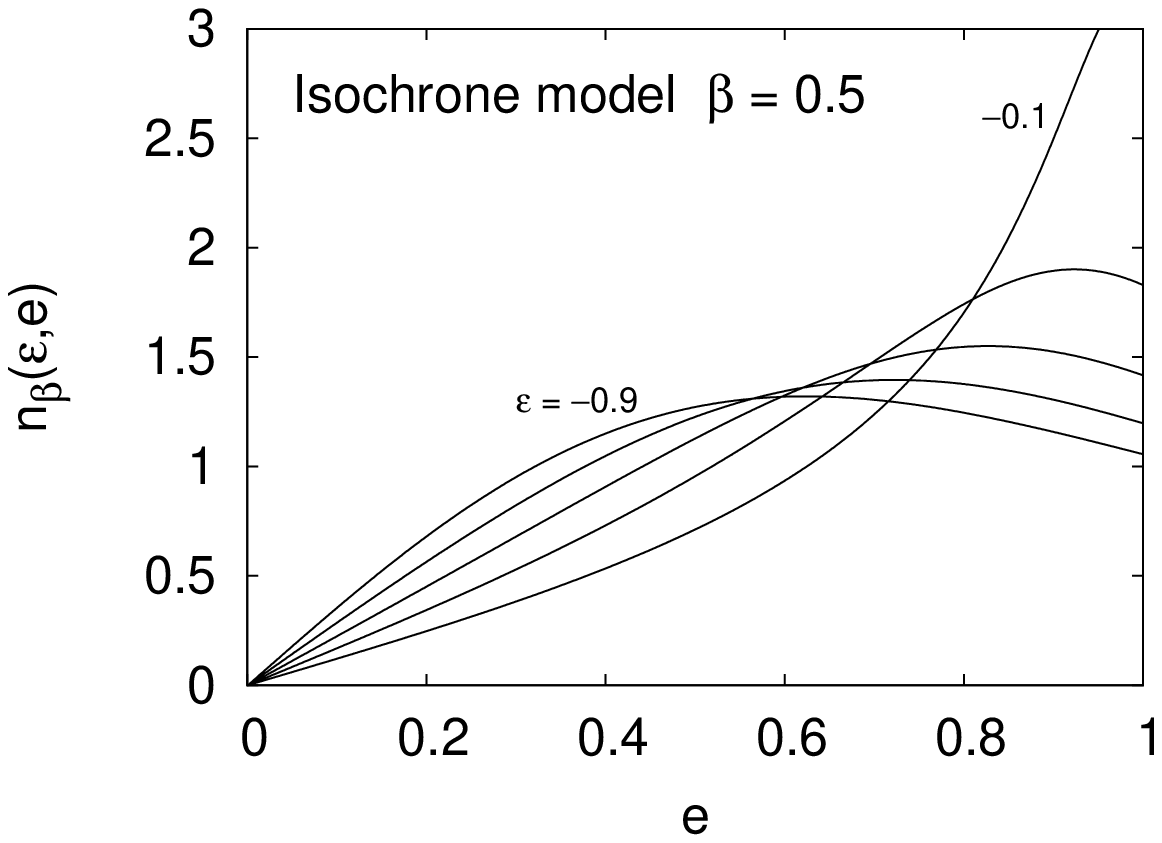}
	\includegraphics[width=\columnwidth]{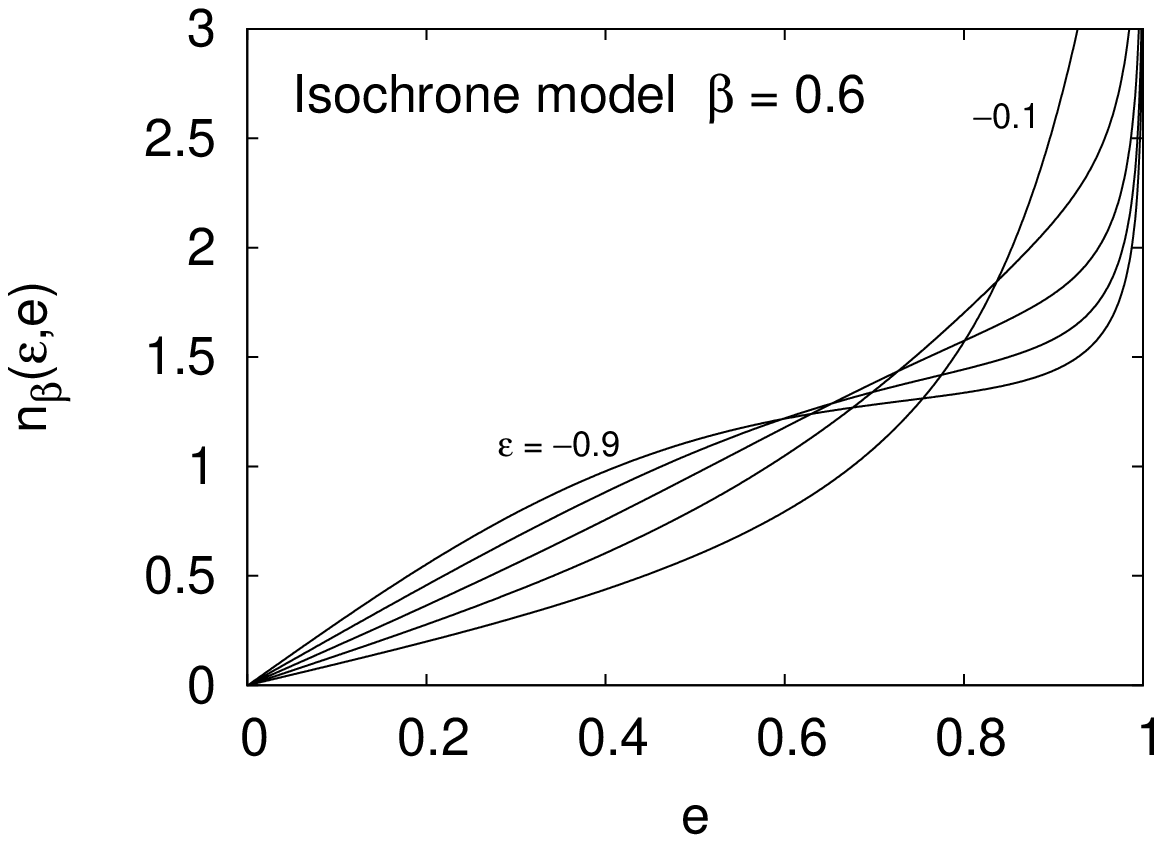}\\
	\includegraphics[width=\columnwidth]{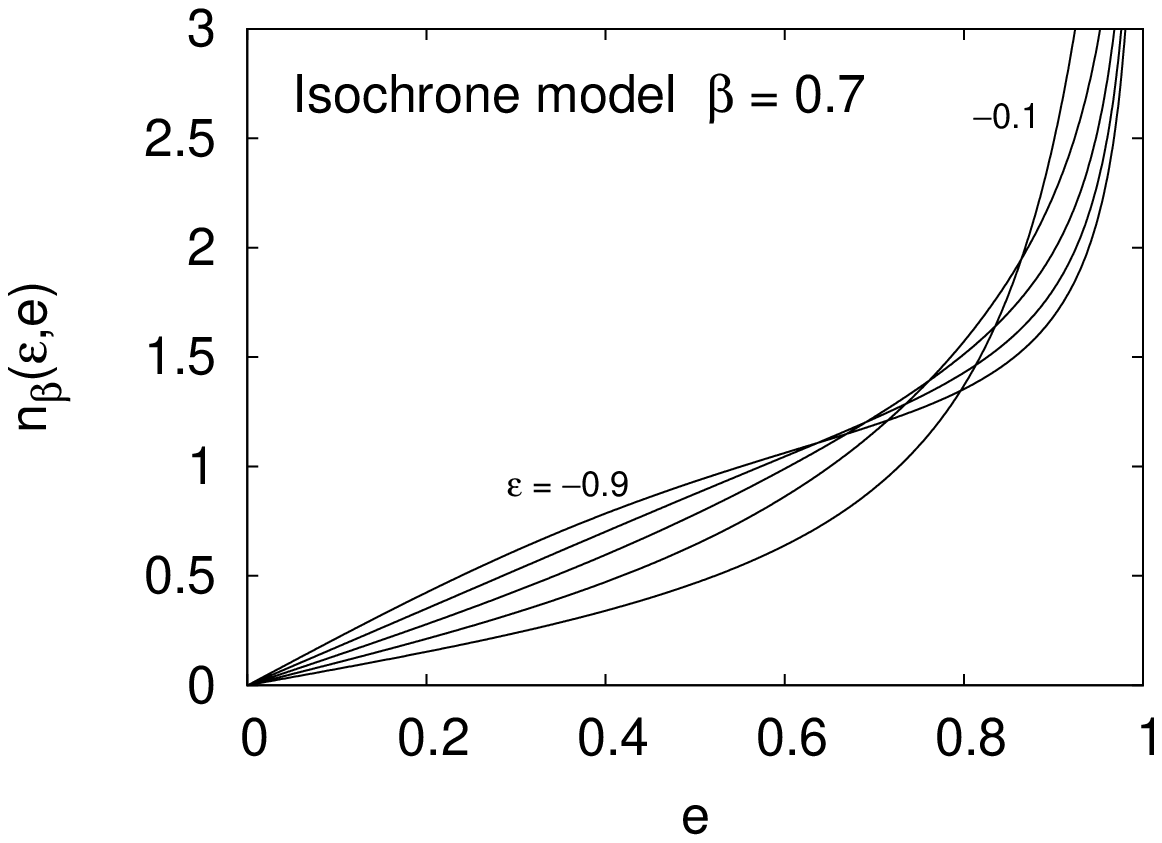}
	\includegraphics[width=\columnwidth]{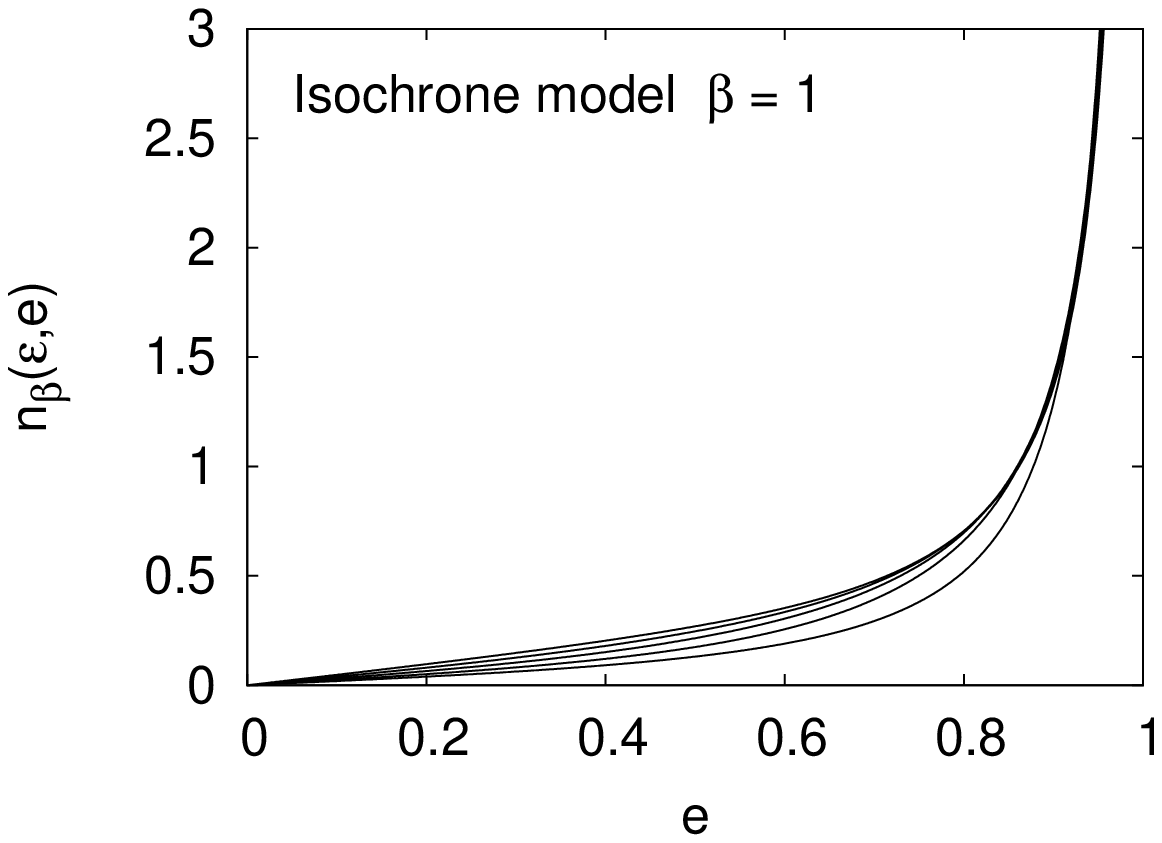}\\
	\caption{Energy-dependent differential distribution of stellar orbital 
eccentricity $n_{\beta}(\varepsilon,e)$ for the isochrone model.  In different 
panels for different values of velocity anisotropy parameter $\beta$, shown 
by lines are the results for dimensionless energy $\varepsilon =-0.9, -0.7, \cdots, -0.1$, 
in steps of $0.2$.  
Note that $n_{\beta}(\varepsilon,e)$ 
is normalized such that $\int_{0}^{1} n_{\beta}(\varepsilon,e)de=1$.  By this normalization, 
the inclination of $n_{\beta}(\varepsilon,e)$ at $e=0$, which is lower for 
smaller $|\varepsilon|$, helps identify each line. } 
\label{Fig isochrone-beta}
\end{center}
\end{figure*}

The shapes of $n_{\beta}(\varepsilon,e)$ for several values of 
$\varepsilon$ and $\beta$ are shown in Figure \ref{Fig isochrone-beta}.  
For any value of $\beta$, there is a general trend such that eccentric 
orbits become more and more dominant as $\varepsilon$ increases.  
However, a marked $\beta$-dependence shows up in the shape of 
$n_{\beta}(\varepsilon,e)$.

When $\beta \lesssim 0.6$, $n_{\beta}(\varepsilon,e)$ has a hump-like 
$e$-distribution with a peak at $e=e_{\rm peak}$.  On the other hand, 
when $0.6 \lesssim \beta \leq 1$, $n_{\beta}(\varepsilon,e)$ has a monotonically 
increasing $e$-distribution with a peak at $e=1$.  In particular, for 
$\beta\approx 0.6$ and $\varepsilon \simeq -1$, $n_{\beta}(\varepsilon,e)$ 
shows something like a trapezoidal shape, similar to the case of 
$\beta\approx 0.6$ for the homogeneous model 
(left panel of Figure \ref{Fig Kepler+IHO-beta}). Furthermore, 
for $\beta > 0.8$, highly eccentric orbits prominently dominate in the 
$e$-distribution.

In order to understand the situation differently, the plots of $e_{\rm peak}$ 
at which the $e$-distribution is peaked for several values of $\varepsilon$ 
and $\beta$ are shown on the left panel of Figure \ref{Fig isochrone+NFW-peak}. 
Here, by taking a 
limit of $\varepsilon$, we can easily confirm, through comparison of this 
figure with Figure \ref{Fig Kepler+IHO-beta}, 
that   
\begin{equation}
\lim_{\varepsilon \to -0} e_{\rm peak}(\beta,\varepsilon) 
= e^{\rm pm}_{\rm peak}(\beta)  ,
\end{equation}
and 
\begin{equation}
\lim_{\varepsilon \to -1} e_{\rm peak}(\beta,\varepsilon) 
= e^{\rm hom}_{\rm peak}(\beta)   ,  
\end{equation}
where superscripts `pm' and `hom' correspond to the point mass model and the homogeneous model, 
respectively. 
More generally, when $\beta>0.6$, we see that $e_{\rm peak}=1$ for any value 
of $\varepsilon$. On the other hand, when $\beta \leq 0.5$, we see that 
$e_{\rm peak}$ is an increasing function of both $\varepsilon$ and $\beta$.  

\subsection{Energy-dependent eccentricity distribution $n_{\beta}(\varepsilon,e)$ for the NFW model} \label{NFW}

Cosmological simulations have been run to reconstruct galaxies from the 
primordial density fluctuations in the universe. These numerical results 
have shown that the dark halo has a universal shape of so-called NFW 
density profile that has little dependence on the cosmology 
(Navarro, Frenk \& White 1997), such as
\begin{equation}
\rho (r) = {\rho}_0 \cdot \frac{a^3}{r{(a+r)}^2}, 
\end{equation} 
where $a$ is the scale length parameter.  This density profile behaves as
$\rho \propto 1/r$ for $r\ll a$, while $\rho \propto 1/r^{3}$ for $r\gg a$.
The associated gravitational potential is of the form 
\begin{equation}
V(r) = -4\pi G {\rho}_0 a^3 \frac{\ln(1+ r/a)}{r} . 
\end{equation} 

Provided $a \neq 0$, we define dimensionless variables and effective 
potential as follows: 
\begin{equation} \label{var_NFW}
x \equiv \frac{r}{a}, 
\hspace{2ex} \varepsilon \equiv \frac{E}{4\pi G \rho_0 a^2}, 
\hspace{2ex} \lambda \equiv \frac{L^2}{4 \pi G \rho_0 a^4}, 
\end{equation}
and 
\begin{equation}
\Phi_{\rm eff}(\lambda;x) \equiv 
\frac{1}{4\pi G \rho_0 a^2} \left[ V(r) + \frac{L^2}{2 r^2} \right]  
\equiv - \frac{\ln(1+x)}{x} + \frac{\lambda}{2 x^2} .  
\end{equation}
Equation (\ref{eq_r1r2}) then reads  
\begin{equation} \label{eqxNFW}
\varepsilon x^2 + x \ln(1+x) - \frac{\lambda}{2} = 0 . 
\end{equation}
This equation indicates a one-to-one correspondence between 
$(\varepsilon,\lambda)$ and $(\varepsilon,e)$, and allows two real 
and positive solutions if and only if 
\begin{equation}
-1 < \varepsilon < 0 \hspace{2ex} {\rm and} \hspace{2ex}
0 < \lambda < \lambda_{\rm cir} \equiv x_{\rm c} \ln (1+x_{\rm c}) - \frac{x_{\rm c}^2}{1+x_{\rm c}}, 
\end{equation}
where $x_{\rm c}$ is the solution for 
\begin{equation}
-2 \varepsilon = \frac{\ln (1+x)}{x} + \frac{1}{1+x} . 
\end{equation} 
We denote the two solutions $x_{\rm peri}$ and $x_{\rm apo}$ 
($x_{\rm apo} > x_{\rm peri}$), and use of them gives 
\begin{equation}
T_r = \sqrt{\frac{1}{2\pi G \rho_0}} 
\int_{x_{\rm peri}}^{x_{\rm apo}} 
\frac{x dx}{\sqrt{\varepsilon x^2 + x\ln(1+x) -\frac{\lambda}{2}}}, 
\end{equation}
\begin{equation}
 L^2 = 8 \pi G \rho_0 a^4 
\left[ \varepsilon x_i^2 + x_i \ln (1+x_i) \right]   
\;\; (x_i= x_{\rm peri} \; {\rm or} \; x_{\rm apo}) ,  
\end{equation}
and  
\begin{multline}
{\left(\frac{\partial L^2}{\partial e}\right)}_{E} = 
- 4 \pi G \rho_0 a^4 
{(x_{\rm apo} + x_{\rm peri})}^2  \\
\times{\left[ 
\frac{x_{\rm apo}}
{x_{\rm peri} \left( 2 \varepsilon + \frac{1}{1+x_{\rm peri}} \right)
+ \ln (1+x_{\rm peri})} 
\right.}
\\
{\left.
- \frac{x_{\rm peri}}
{x_{\rm apo} \left( 2 \varepsilon + \frac{1}{1+x_{\rm apo}} \right)
+ \ln (1+x_{\rm apo})} 
\right]}^{-1} .
\end{multline}
We see that $n_{\beta}(\varepsilon,e)$ does not allow analytic 
expression in terms of $\varepsilon$ and $\beta$.  Accordingly, 
derivation of $n_{\beta}(\varepsilon,e)$, as well as $N_{\beta}(e)$
with the weight function $g(\varepsilon)$, needs full numerical 
integration for the NFW model. 

\begin{figure*}
	\begin{center}
	\includegraphics[width=\columnwidth]{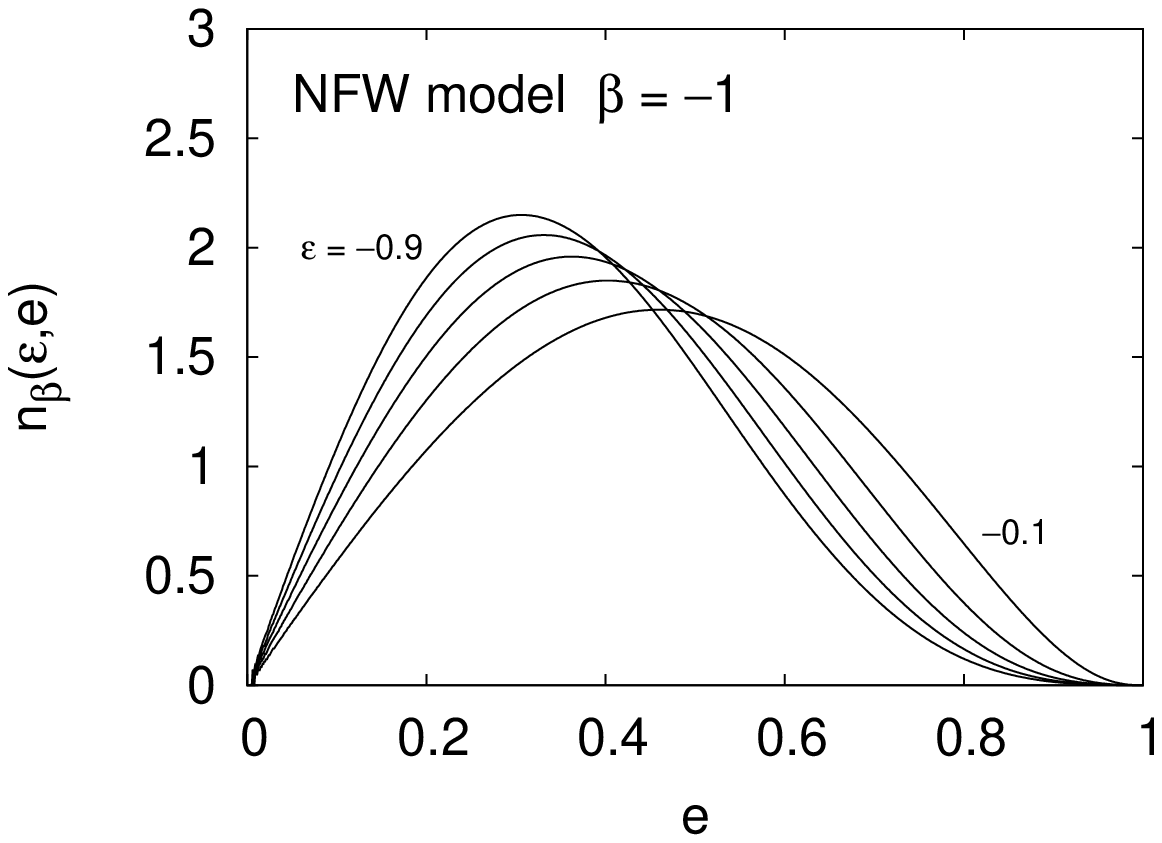}
	\includegraphics[width=\columnwidth]{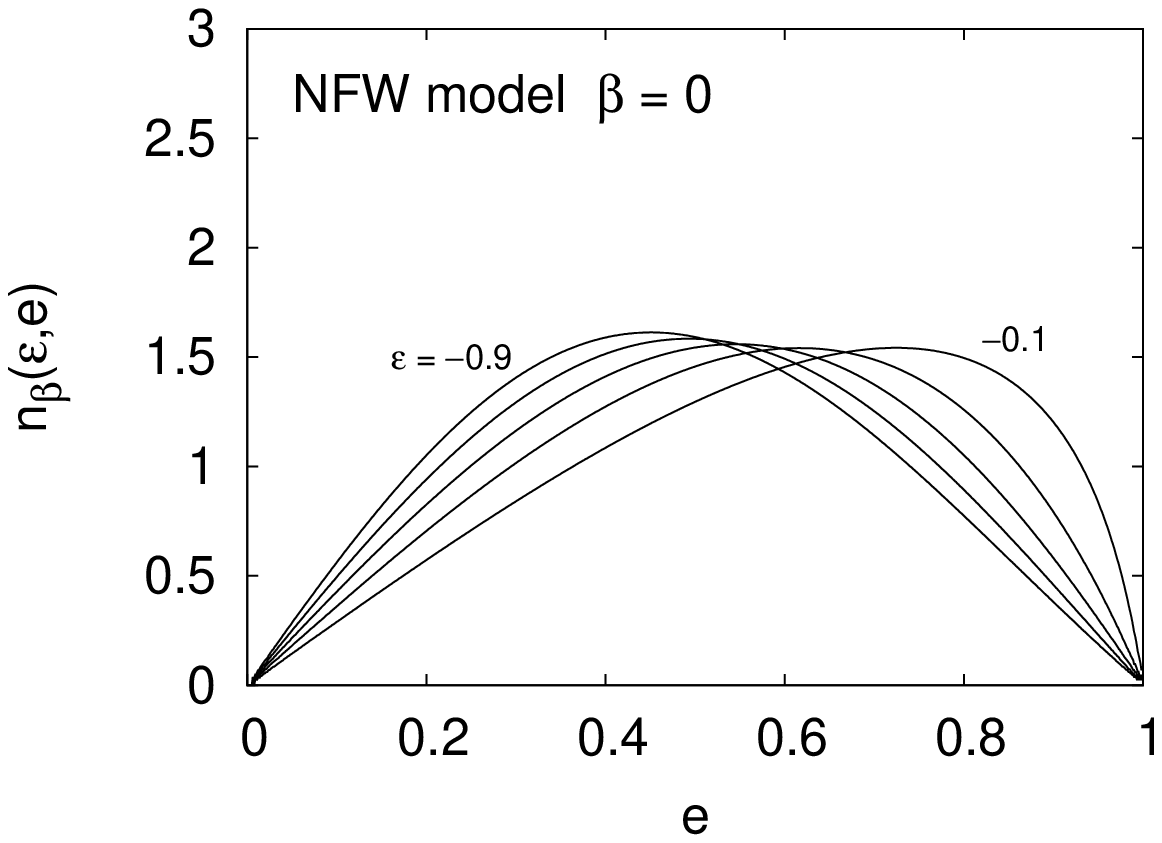}\\
	\includegraphics[width=\columnwidth]{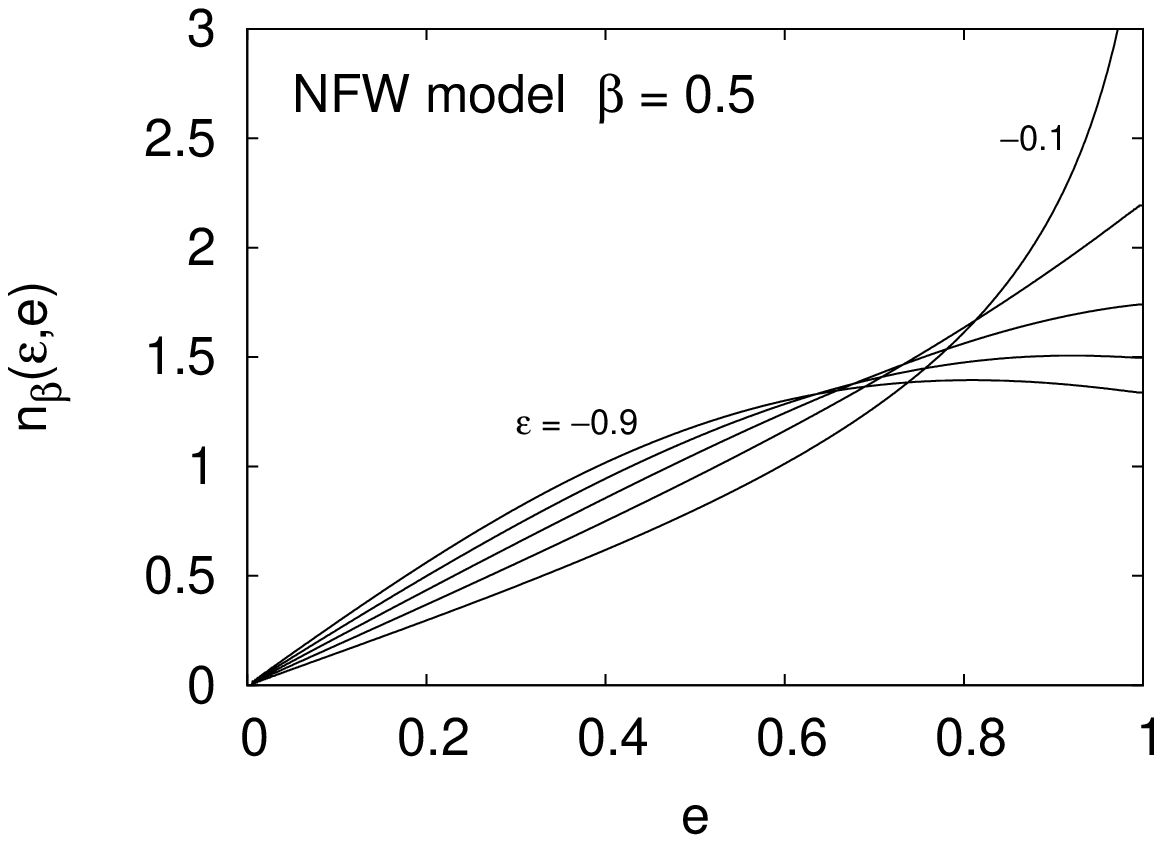}
	\includegraphics[width=\columnwidth]{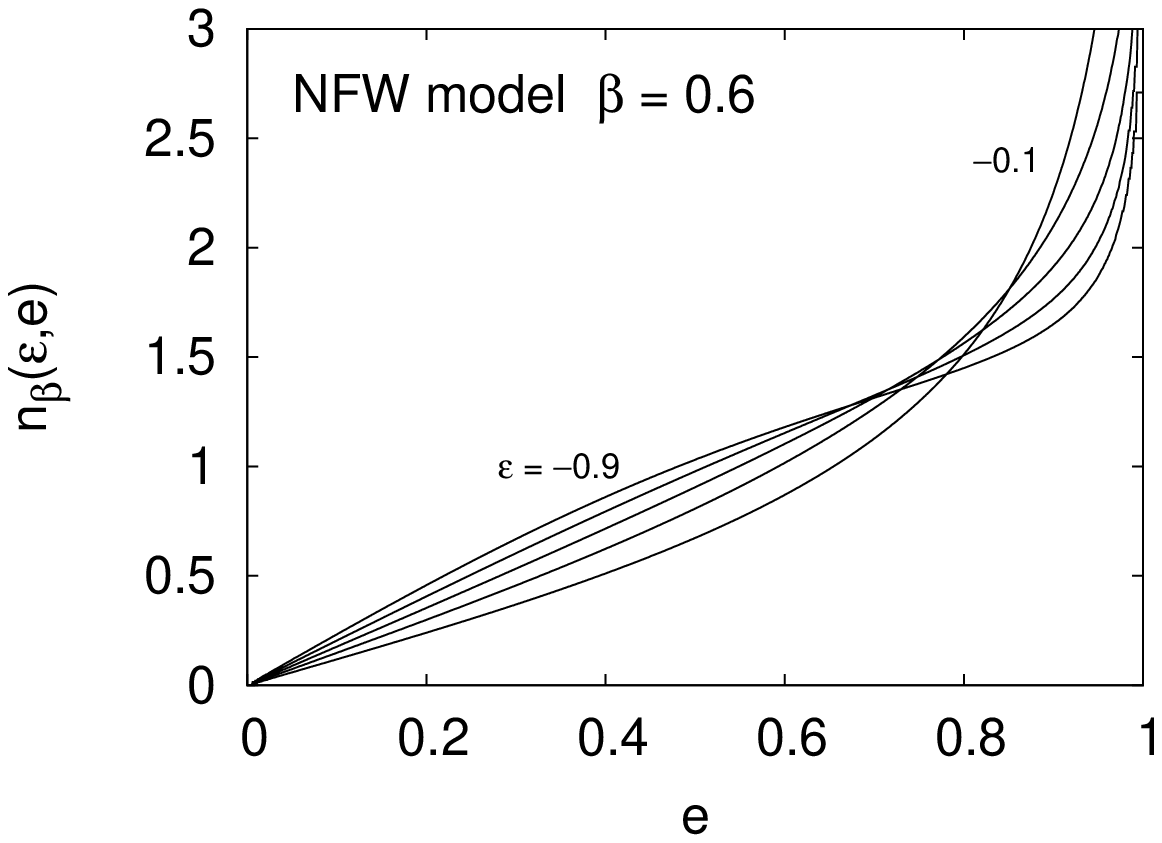}\\
	\includegraphics[width=\columnwidth]{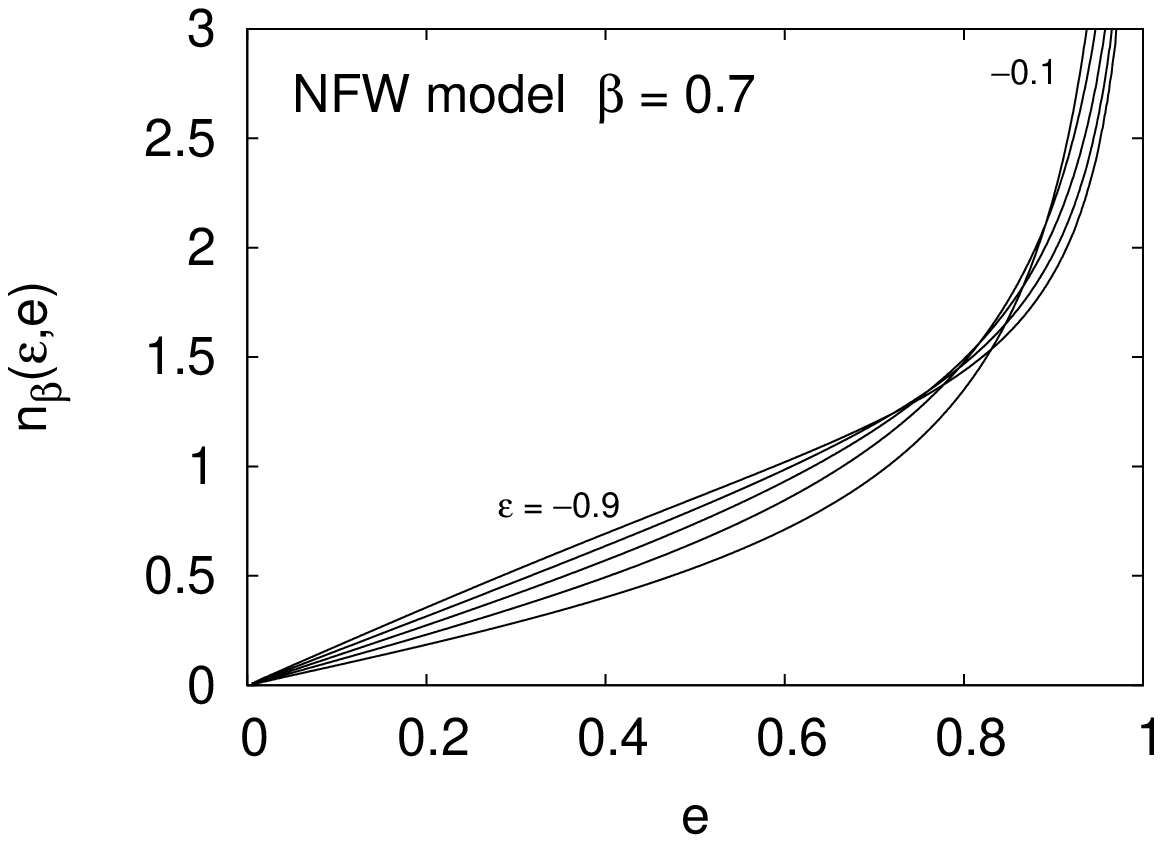}
	\includegraphics[width=\columnwidth]{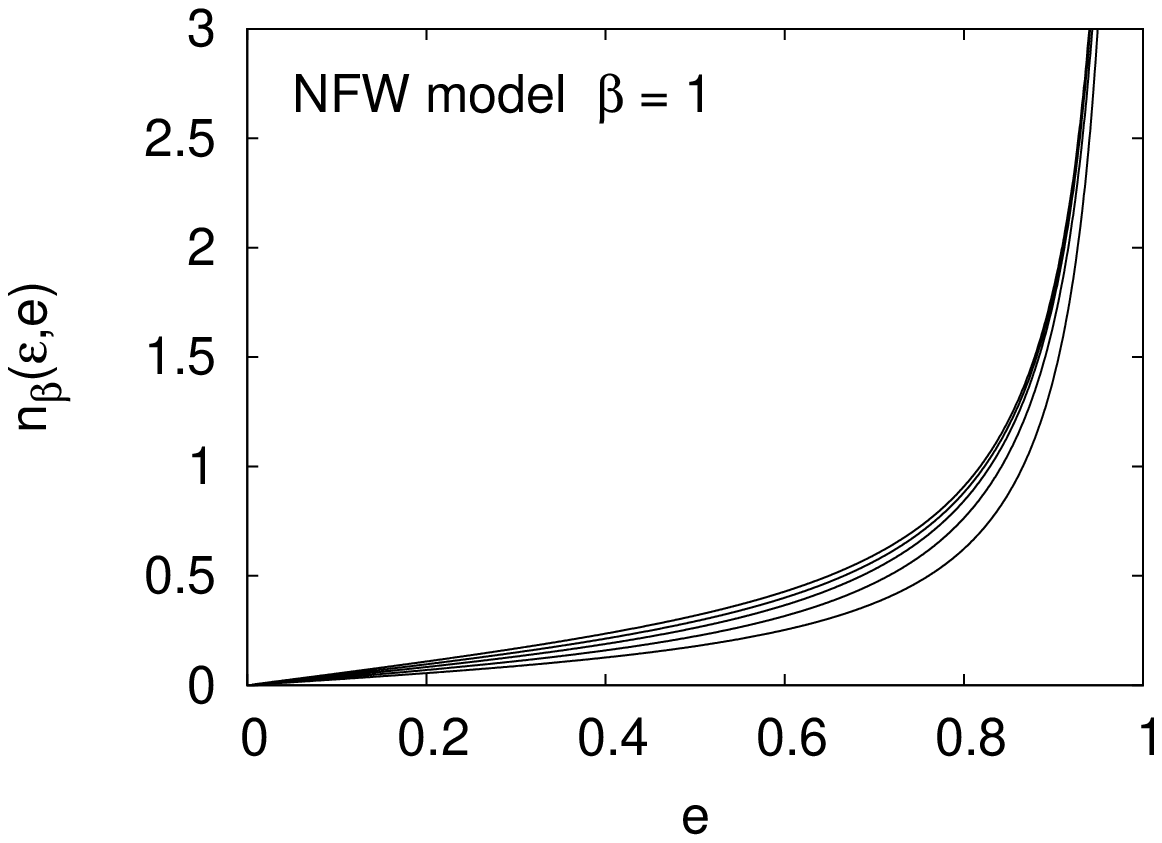}\\
	\caption{Energy-dependent differential distribution of stellar orbital 
eccentricity $n_{\beta}(\varepsilon,e)$ for the NFW model.  Others are the same as 
in Figure \ref{Fig isochrone-beta}.} 
\label{Fig NFW-beta}
\end{center}
\end{figure*}

\begin{figure*}
\begin{center}
	\includegraphics[width=\columnwidth]{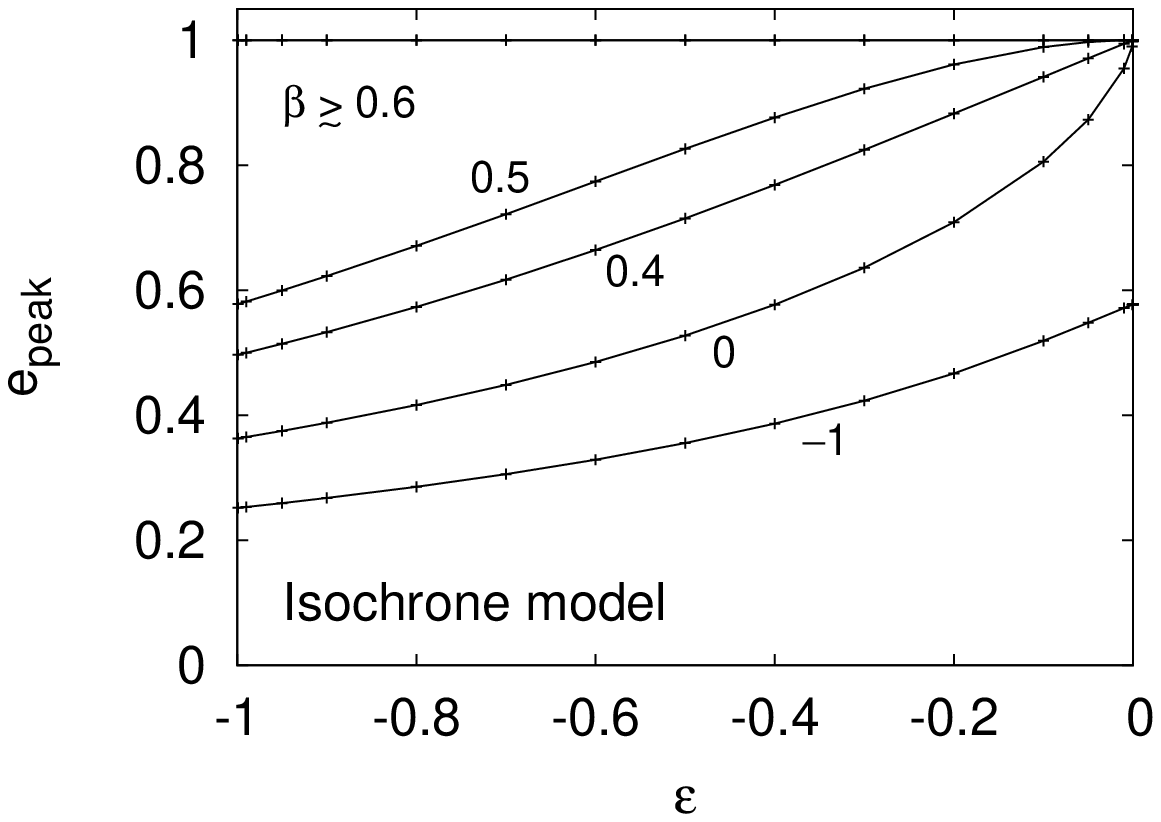}
        \includegraphics[width=\columnwidth]{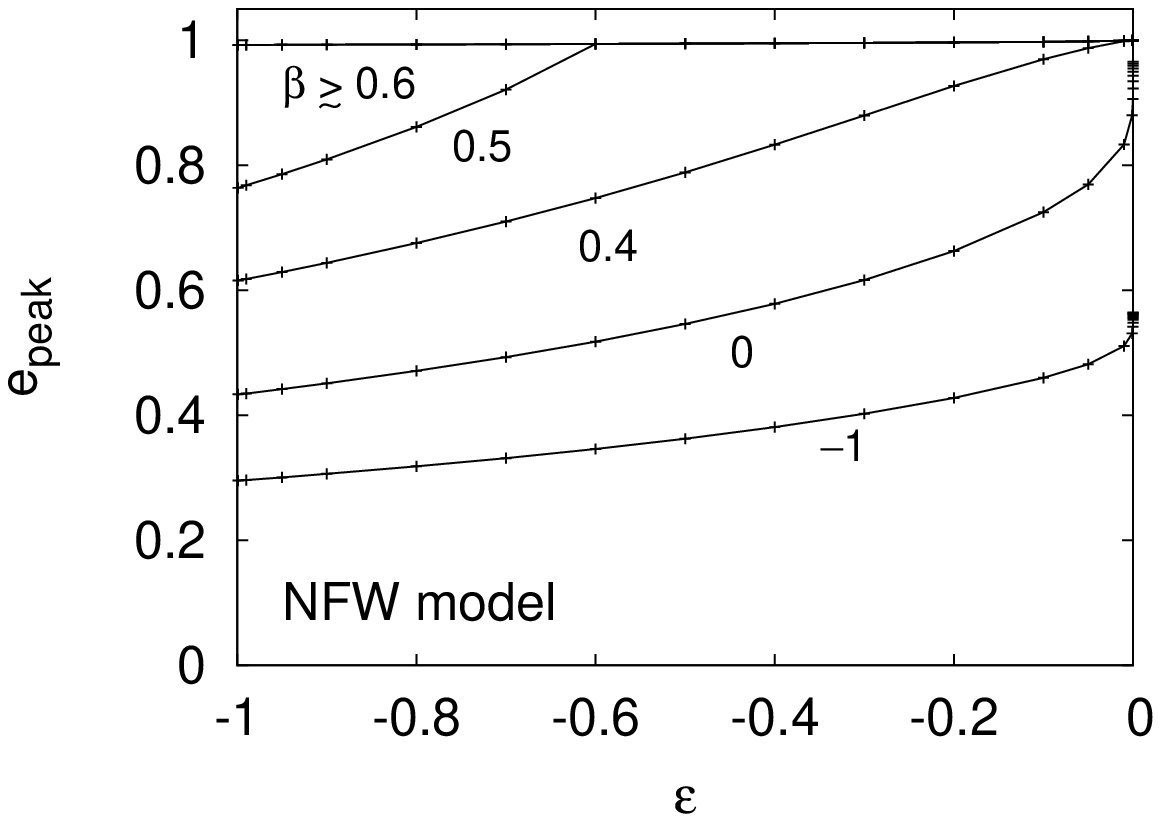}\\
	\caption{The peak eccentricity $e_{\rm peak}$ as a function of dimensionless  
energy $\varepsilon$ for the isochrone model (left panel) and the NFW model (right panel).
The results are shown by lines for different values of velocity anisotropy parameter $\beta$.}
\label{Fig isochrone+NFW-peak}
\end{center}
\end{figure*}

The results of $n_{\beta}(\varepsilon,e)$ for several values of 
$\varepsilon$ and $\beta$ are shown in Figure \ref{Fig NFW-beta}. 
The plots of $e_{\rm peak}$ at which the $e$-distribution is peaked 
for several values of $\varepsilon$ and $\beta$ are shown on the 
right panel of Figure \ref{Fig isochrone+NFW-peak}.  
Here, similarly to the isochrone model, 
by taking a limit of $\varepsilon$, we can easily confirm that
\begin{equation}
\lim_{\varepsilon \to -0} e_{\rm peak}(\beta,\varepsilon) 
= e^{\rm pm}_{\rm peak}(\beta) ,
\end{equation}
and 
\begin{equation}
\lim_{\varepsilon \to -1} e_{\rm peak}(\beta,\varepsilon) 
= e^{\rm lp}_{\rm peak}(\beta)  , 
\end{equation}
where superscripts `pm' and `lp' correspond to the point mass model and 
the linear potential model described in Appendix B.1, respectively. 

Except for slight shift of the $e$-distribution to have more weight 
at higher $e$, overall behavior of $n_{\beta}(\varepsilon,e)$ for 
the NFW model is very similar to the isochrone model.  Such slight 
shift occurs, because the mass is little more centrally concentrated 
in the NFW model compared with the isochrone model.  

The insensitivity to the choice of gravitational potential, as far 
as it remains realistic, is encouraging, especially when our 
theoretical $e$-distribution is to be compared with that observed 
for stars in the Milky Way halo. 

\subsection{Results of $N_{\beta}(e)$} \label{result}

In the previous subsections, we have derived the $E$-dependent form 
of $n_{\beta}(E,e)$ for the respective models of isochrone and NFW. 
In order to obtain their eccentricity distribution $N_{\beta}(e)$ in 
equation (\ref{general-N}), we have to specify the weight function of 
$g(E)$ which can in principle be derived in a self-consistent way 
(Lynden-Bell 1962, 1963). Here, instead of entering into robustness, 
however, we take a simple approximation of $g(E)$ as having 
the form:
\begin{equation}
  g(E) = A \exp \left( - \frac{E}{\sigma^2} \right) ,
\end{equation}
where $A$ is a constant and $\sigma$ stands for the radial velocity 
dispersion $\sigma_r\sim 150\;{\rm km\;s^{-1}}$ 
for the Milky Way halo stars 
(e.g. Yoshii \& Saio 1979; Chiba \& Beers 2000, 2001). 

We can imagine that halo stars traveling far distantly from the 
galaxy center with near-zero energy would be captured by adjacent 
dark halo.  
Thus, it is reasonable to introduce a truncation energy $E_{\rm t}$ 
above which $g(E)$ should vanish. 

The NFW model provides a direct reason to include $E_{\rm t}$ in 
the analysis.  The mass of dark halo within the radius $r$ is 
naively given by 
\begin{equation}
   M(r) = 4\pi \rho_0 a^3 \left[ \ln\left( 1+\frac{r}{a} \right) - \frac{r/a}{1 + r/a} \right] 
\end{equation}
and diverges in the limit of large $r$.  In fact, numerical simulations 
indicate that the NFW density profile applies only inside a certain boundary 
radius but does not apply beyond it because of the existence of adjacent 
dark halos.  Such a boundary usually used is the virial radius $r_{200}$ 
within which the averaged density is equal to $200$ times the critical 
density of the universe and the effects by adjacent dark halos are negligible.  
Thus, it is reasonable to place $E_{\rm t}$ at $V(r_{200})$ and assume that while 
halo stars with $E<E_{\rm t}$ stay in the system, those with $E>E_{\rm t}$ could be unbound 
and leave the system. 

From all these considerations, we examine how $N_{\beta}(e)$ would be 
modified with $E_{\rm t}$ taken into account in the analysis. 
Here, we set $E_{\rm t}$ equal to the potential energy $V(r_{200})$ 
and write it in the dimensionless form: 
\begin{equation}
    \varepsilon_{\rm t} = - \frac{\ln(1+c)}{c}, 
\end{equation}
where $c$ is the concentration parameter defined as $c\equiv r_{200}/a$.  
Use of the kinematic data of the blue horizontal branch stars in the Milky Way halo 
and some CDM simulations of a halo of $M(r_{200}) \sim 10^{12}M_\odot$ as massive as  
the Milky Way halo gives $c=3.9-12.5$ (Xue et al. 2008), which corresponds to 
$\varepsilon_{\rm t} = -0.4$ to $-0.2$.    
Thus, a choice of this range of $\varepsilon_{\rm t}$, together with $\beta=0.5-0.7$ 
(cf. section 2.2),  would be appropriate for our analysis of the Milky Way halo.

We have repeated the calculations of $N_{\beta}(e)$ for several values of 
$\sigma$ and $\varepsilon_{\rm t}$ in the integration of $n_{\beta}(\varepsilon,e)$ 
over $\varepsilon$ in equation (\ref{general-N}), and find that $N_ {\beta}(e)$ 
is insensitive to $\sigma$ but sensitive to $\varepsilon_{\rm t}$.  The shape of 
$N_{\beta}(e)$ is almost the same as that of $n_{\beta}(\varepsilon_{\rm t},e)$. 
This is because $n_{\beta}(\varepsilon_{\rm t},e)$ significantly contributes to the 
integration of $n_{\beta}(\varepsilon, e)$.  For example, in a particular 
case of $\beta=0$ for the isochrone model, we clearly see such a situation 
from the explicit expression:  
\begin{equation}
  \int_{0}^{1} n_{\beta = 0}(\varepsilon,e) de 
  = 8 \pi^3 \sqrt{GM b^5} {(1+\varepsilon)}^2 {(-\varepsilon)}^{-\frac{5}{2}} . 
\end{equation} 

\begin{figure*}
\begin{center}
	\includegraphics[width=\columnwidth]{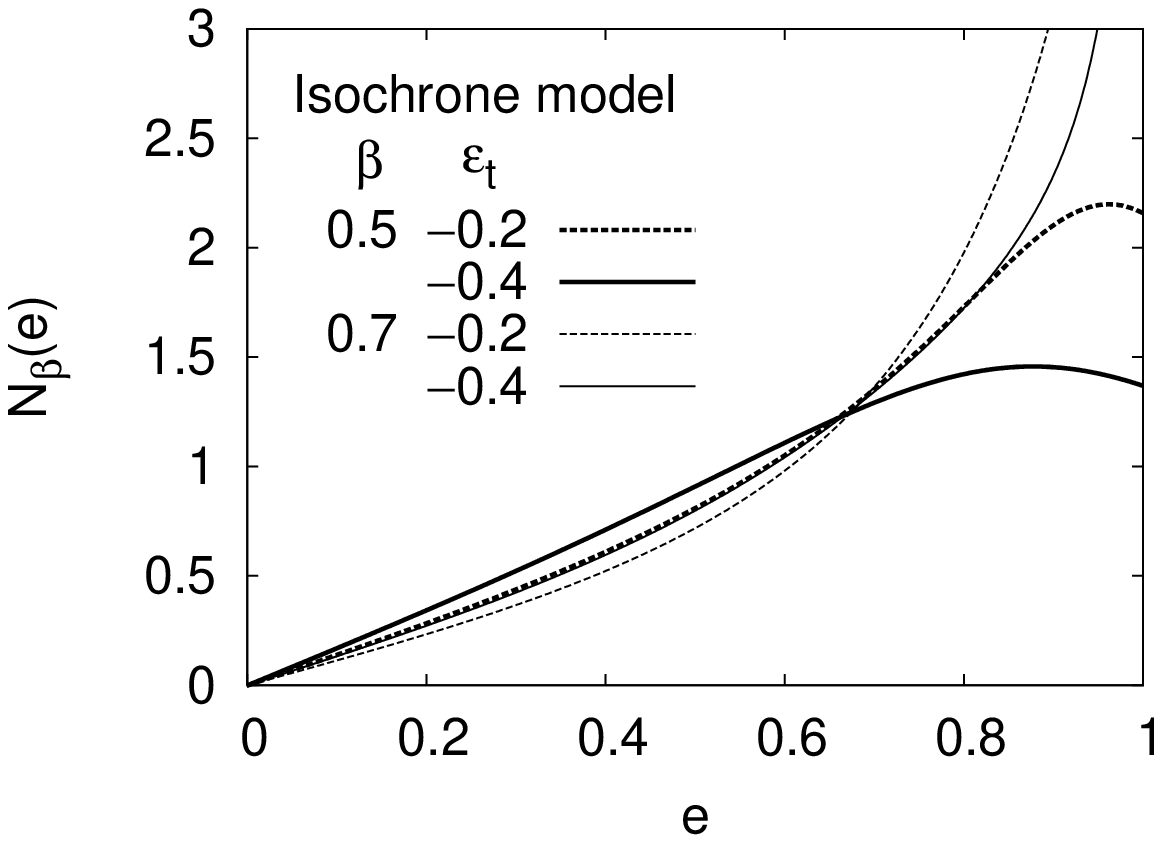}
	\includegraphics[width=\columnwidth]{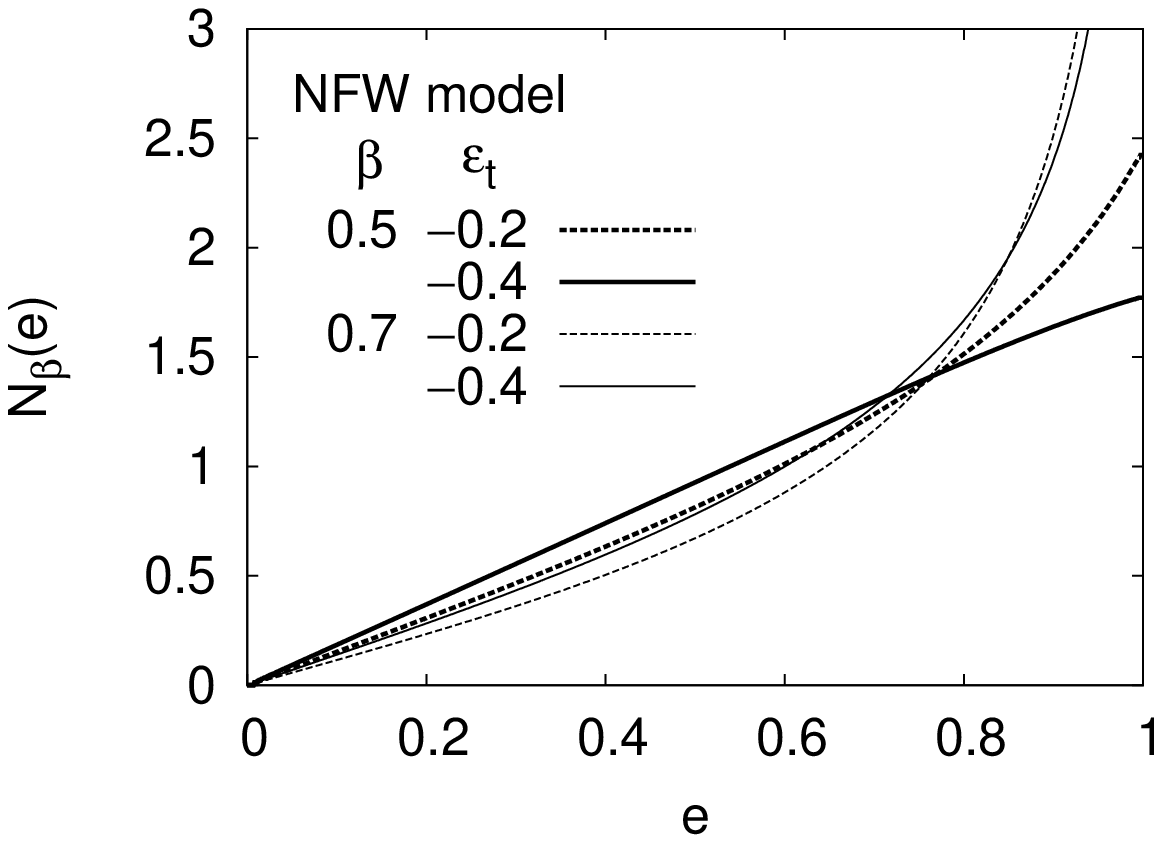}\\
	\caption{Differential distribution of stellar orbital eccentricity $N_{\beta}(e)$ 
for the isochrone model (left panel) and the NFW model (right panel). Adopted combinations 
of $(\beta,\varepsilon_{\rm t})$ more or less agree with observations of the Milky Way halo. 
Note that the normalization factor of $N_{\beta}(e)$ is arbitrarily chosen so that the nearly 
linear trend up to $e\approx 0.7$ is clearly seen. } 
\label{Fig isochrone+NFW-N}
\end{center}
\end{figure*}

Using the typical combinations of $(\beta,\varepsilon_{\rm t})=(0.5, -0.2)$, 
$(0.5, -0.4)$, $(0.7, -0.2)$, and $(0.7, -0.4)$ that more or less 
agree with observations of the Milky Way halo, the results of 
$N_{\beta}(e)$ for both the isochrone and NFW models are shown in 
Figure \ref{Fig isochrone+NFW-N}.  We see from this figure that as far 
as reasonable values of $\beta$ and $\varepsilon_{\rm t}$ are adopted, the 
resulting shape of $N_{\beta}(e)$ should be almost linearly proportional 
to $e$, except for the deviation only at $e>0.7$.  This is largely 
regardless of adopting either the isochrone model or the NFW model. 
Thus, if the dominant component of the Milky Way halo is in dynamical equilibrium, 
the total eccentricity distribution of stellar halo 
is expected to have a linear trend at $e<0.7$ similar to our results. 
On the other hand, the behavior of predicted $N(e)$ at $e>0.7$, which still 
shows little difference between the isochrone and NFW models, is sensitive 
to $\beta$ and $\epsilon_{\rm t}$. Consequently, such sensitivity can be 
used for a consistency check of the assumed form of $f(E,L)$.  These  
predictions in the separate regions of $e<0.7$ and $e>0.7$ are testable, 
given that large kinematical data of halo stars are available at present from 
the SEGUE project or in the near future from the Gaia project.

\section{Summary and discussion} \label{discussion}

Hierarchical clustering scenarios of galaxy formation suggest that the major  
merger of at least several subhalos with comparable masses would occur at the 
last stage of galaxy formation.  This last major merger would cause the violent 
relaxation of halo stars and make them in dynamical equilibrium with a dark halo.  
Based on the assumptions that approximate such a status just after the last 
violent relaxation (section \ref{intro}), we have presented theoretical 
predictions of $N(e)$ for halo stars. 
This predicted $N(e)$ should be observed for the Milky Way halo 
if it is an isolated system and the subsequent variation of the potential 
is quiescent enough to conserve the eccentricity of each star. 

However, recent nearby observations suggest that 
at least some part of the Milky Way halo may have originated from accreted satellites, 
which possibly deviates the observed $N(e)$ from our predictions. 
For example, 
if infalling satellites break up and spread their member stars into the field, 
these stars would show peculiar eccentricity distribution which necessarily 
imprints the initial condition of the progenitor satellites. 
In addition, 
if such satellites locally disturb the halo potential, 
some {\it in-situ} halo stars may have altered their orbits 
(e.g. Zolotov et al. 2009). 
With an invention of segregating  {\it in-situ} 
halo stars from infalling stars, we might be able to well understand 
the nature of accretion and distortion of satellites.

Numerous authors subdivided halo stars into some `components' and examined the correlations between chemistry, age and kinematics of stars in each component.  
Carollo et al. (2010) obtained reliable eccentricities 
for $\sim 10,000$ halo stars within $4\;{\rm kpc}$ of the sun 
and decomposed them into the inner and outer halo components having distinct  eccentricity distributions from each other. 
Since their sample is {\it local} and is inherently biased in favour of stars that stay
longer in the surveyed region, our formalism, which is designed to predict
$N(e)$ of the whole stellar halo, has to be modified for the purpose of fair comparison
with their data. Through proper incorporation of effects of such a bias, we can still
predict $N(e)$ for a {\it local} sample by fully taking into account a probability of
finding each of halo stars in the surveyed region. This will be done in a separate paper
in preparation. On the other hand, our formalism can directly apply to a {\it global},
and therefore less biased, sample of halo stars with reliable orbital eccentricities,
such as those from next generation surveys including the Gaia mission. 
In either case, the analytical approach in the present paper certainly forms a 
basis that serves as a useful tool for analysing the kinematics of the stellar halo. 

Large, unbiased database of halo stars 
would enable us to test whether a given component is in dynamical equilibrium 
by comparing the observed and predicted shape of $N(e)$. 
Such comparison would hopefully discover some relaxed components, and their adiabatically conserved shape of $N(e)$ would carry some useful information of the physics of violent relaxation. 
Moreover, the spatial distribution of these relaxed components would enable us to 
see how far out 
in the halo the violent relaxation has exerted and how strongly it has affected 
the stellar halo.  If the information of last violent relaxation, yet to be known observationally, 
is gained in this way, more precise assessment to the early evolution of the Milky 
Way would be possible, and our understanding of its formation would greatly be advanced. 

Our current calculations of $N(e)$ are certainly very simple and can be improved 
by using more realistic assumptions. 
For example, we can modify our analysis to allow axisymmetric potentials 
including a disk-like component as well as a bulge. 
Preliminary analysis has confirmed that inclusion of a disk-like component would 
cause no significant change in the linear trend of $N(e)$ described in section \ref{result}, 
which will be discussed in a separate paper. 
Also, our choice of $f(E,L)$ having the form in equation (\ref{f_beta}) has to be extended to allow the radial 
dependence of $\beta(r)$. 
Further elaborate modeling of $N(e)$ with these theoretical improvements, when 
applied to future large survey of halo stars, would then provide a promising way 
of unraveling mysteries of the galaxy formation and  evolution in a paradigm of hierarchical clustering in the $\Lambda$CDM cosmology.

\section*{Acknowledgments}

We thank Beers, T., Carollo, D., Minezaki, T., Tsujimoto, T., and Yamagata, T. for useful discussions 
and suggestions.

\appendix

\section{Allowed region of $(E,L)$ for bound orbit} \label{allowed}

A steady, bound orbit in a gravitational potential $V(r)$ 
generated by a density distribution $\rho (r)$ is only 
possible in a subset of energy $E$ and angular momentum $L$ that 
allows two real and positive solutions for equation (\ref{eq_r1r2}).  
We discuss such an allowed region of $(E,L)$ in this appendix.  

We begin with the effective potential 
\begin{equation}
V_{\rm eff}(L;r) = V(r) + \frac{L^2}{2r^2} . 
\end{equation}
Then, from the  definition, we obtain 
\begin{equation}
{\left( \frac{\partial}{\partial r} V_{\rm eff}(L;r) \right)}_L = 
\frac{1}{r^3}  \left[ G M(r) r - L^2 \right] , 
\end{equation}
where $M(r)$ is the total mass inside the radius $r$.  Since $GM(r)r$ 
is a monotonically increasing function of $r$ and it satisfies 
\begin{equation}
   \lim_{r  \to 0} \left[ G M(r) r \right] = 0 ,
\hspace{2ex} {\rm and} \hspace{2ex}
\lim_{r \to \infty} \left[ G M(r) r \right] = \infty ,
\end{equation}
there always exists an radius $r_{\rm c} = r_{\rm c}(L)$ for which $GM(r_{\rm c})r_{\rm c} = L^2$, 
thus yielding
\begin{equation}
{\left( \frac {\partial}{\partial r} V_{\rm eff}(L;r) \right)}_L 
\grole 0, \;\; \text{if $r \grole r_{\rm c}$} . 
\end{equation}
Since we have 
\begin{equation}
\frac{d}{d(L^2)} r_{\rm c} = 
{\left[ G \left( 4 \pi r_{\rm c}^3 \rho(r_{\rm c}) + M(r_{\rm c}) \right) \right]}^{-1} > 0  ,
\end{equation}
and 
\begin {equation}
\frac{d}{d(L^2)} V_{\rm eff}\left( L;r_{\rm c}(L) \right) = 
\frac{1}{2 r_{\rm c}^2} > 0 , 
\end{equation}
the allowed range of $E$ with $L$ fixed can be expressed as 
\begin{equation} \label{allowed-E}
V_{\rm eff}\left( L;r_{\rm c}(L) \right) < E < 0. 
\end {equation}
Here, we define the zero of $V(r)$ so that $\lim_{r \to \infty} V(r) = 0$. 
Thus, for any given $L$, we obtain 
\begin{equation}
\lim_{r \to \infty} V_{\rm eff}(L;r) = 0, 
\end{equation}
which validates that the upper bound of inequality (\ref{allowed-E}) should 
be zero.  As for the allowed region of $L$ when $E$ is fixed, we obtain 
\begin{equation}
0 < L <  L_{\rm cir}(E), 
\end{equation}
for which $L_{\rm cir}(E)$ is the solution of 
\begin{equation}
E = V_{\rm eff}\left( L_{\rm cir};r_{\rm c}(L_{\rm cir}) \right) . 
\end{equation}

\section{Other models of truncated mass distribution}

We present the derivation of $N_{\beta}(e)$ for two models 
of truncated power-law mass distribution: 
\begin{equation} \label{power-law}
    \rho(r) = 
    \begin{cases}
        \frac{(3-\gamma) M}{4 \pi r_{\rm t}^3} {\left( \frac{r}{r_{\rm t}} \right)}^{-\gamma} ,\;\; ({\gamma < 3}) & \text{if $r < r_{\rm t}$} \\
        0 ,                       & \text{otherwise,} 
    \end{cases}
\end{equation}
where $M$ is the total mass of the dark halo and $r_{\rm t}$ is the 
truncation radius.  We note that the truncated homogeneous model 
presented in section \ref{homo} is a special case of $\gamma = 0$ in 
equation (\ref{power-law}). 

\subsection{Linear potential model ($\gamma = 1$)}

\begin{figure*}
\begin{center}
	\includegraphics[width=\columnwidth]{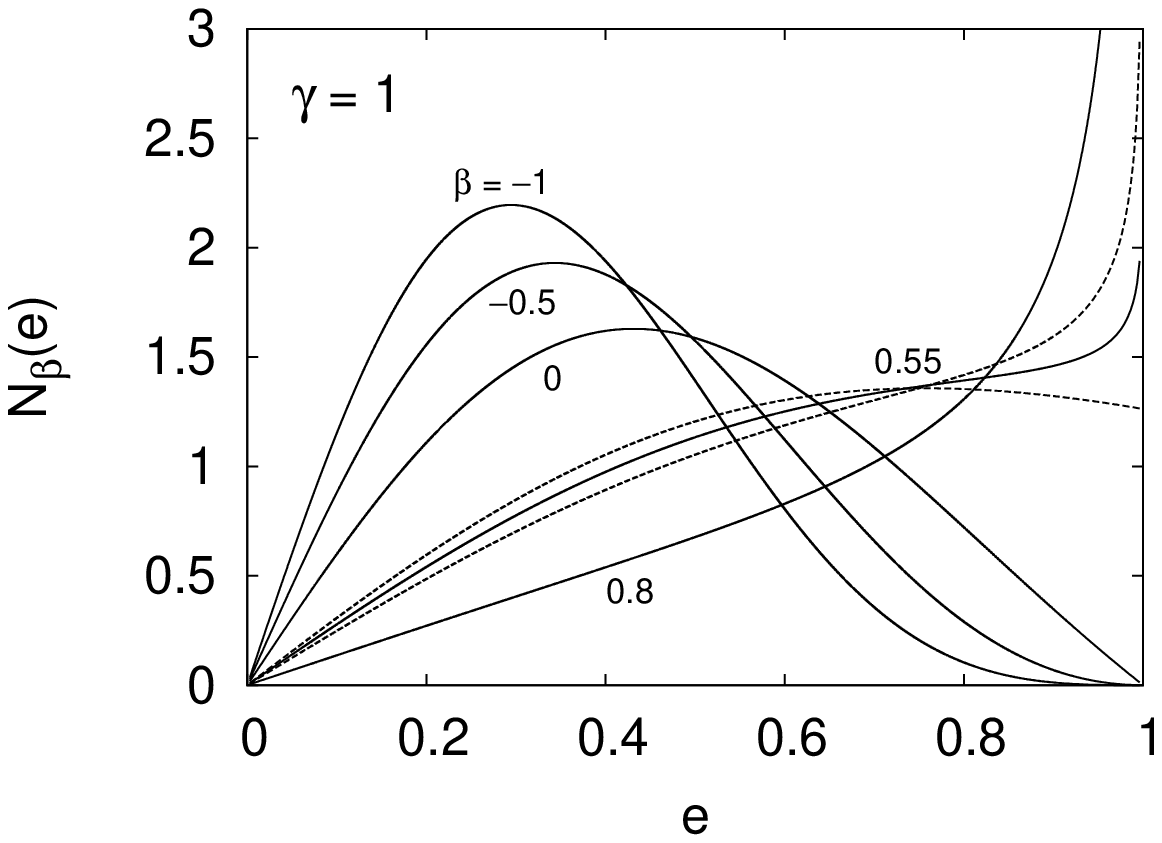}
	\includegraphics[width=\columnwidth]{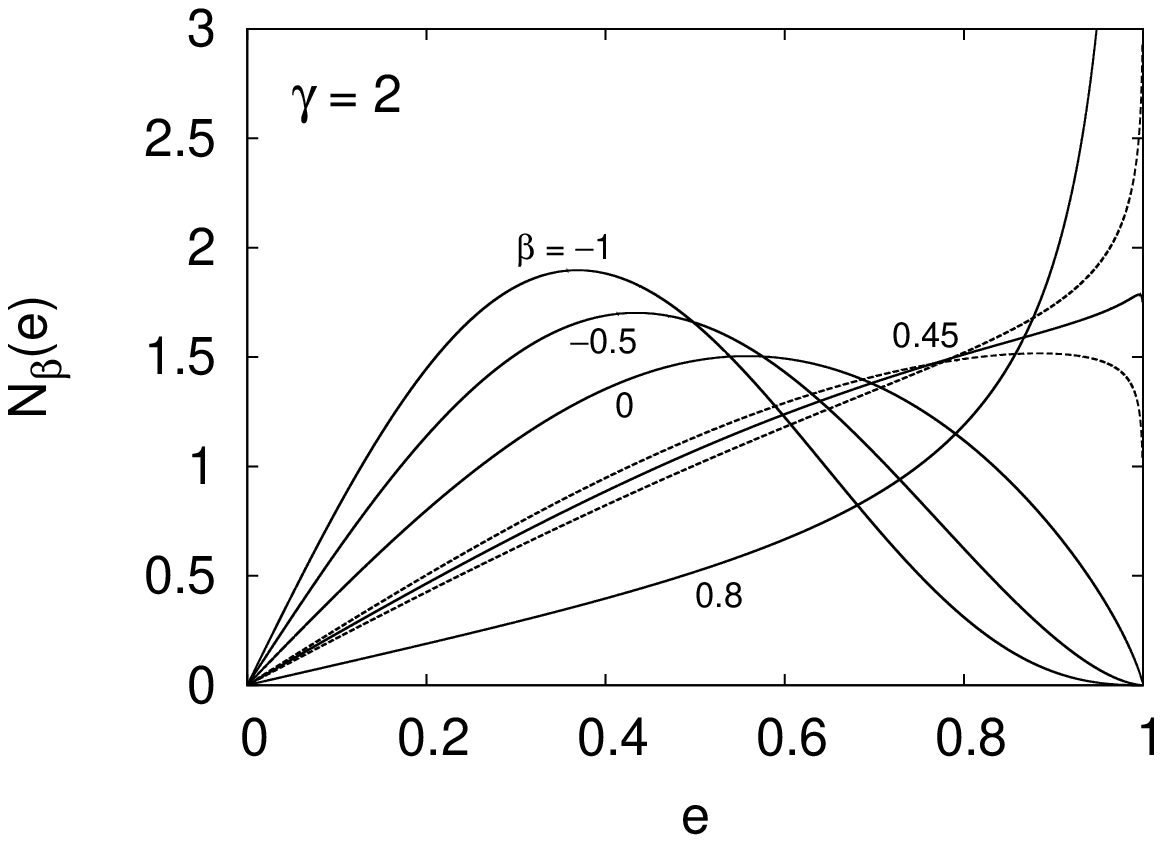}\\
	\caption{Differential distribution of stellar orbital eccentricity $N_{\beta}(e)$
in two cases of truncated mass distribution, such as the linear potential model ($\gamma =1$) 
on the left panel and the singular isothermal model ($\gamma =2$) on the right panel. 
The results are shown by lines for several values of velocity anisotropy parameter $\beta$.  
If $N_{\beta}(e)$ near $e=1$ sensitively changes at some particular value of $\beta$, 
the results for $\beta\pm 0.05$ are additionally shown by dotted lines for the 
purpose of illustrating its sensitivity.  
Note that $N_{\beta}(e)$ is normalized such that $\int_{0}^{1} N_{\beta}(e)+de=1$.} 
\label{Fig linear+SIS-beta}
\end{center}
\end{figure*}

The NFW density profile has a central cusp and behaves like  
$\rho(r) \propto 1/r$ in the limit of Small $r$. This density 
profile corresponds to $\gamma = 1$ in equation (\ref{power-law}), 
and we have  
\begin{equation}
    \rho(r) = 
    \begin{cases}
        \frac{M}{2 \pi r_{\rm t}^3} {\left( \frac{r}{r_{\rm t}} \right)}^{-1} , & \text{if $r < r_{\rm t}$} \\
        0 ,                       & \text{otherwise.} 
    \end{cases}
\end{equation}
The gravitational potential arising from this density profile 
is given by 
\begin{equation}
V(r) = 
\begin{cases}
     - \frac{2GM}{r_{\rm t}} + \frac{GM}{r_{\rm t}} \left( \frac{r}{r_{\rm t}}\right) , &\text{if $r < r_{\rm t}$}\\
     -\frac{GM}{r} ,  & \text{otherwise,}
\end{cases}
\end{equation}
and we will refer to this potential a `truncated linear potential.' 
We consider only stars with $E<E_{\rm t}\equiv-GM/r_{\rm t}$, 
which guarantees the stars to be confined inside the truncated radius $r_{\rm t}$.  
Thus, bound orbits within the truncated sphere are allowed if 
$E_{\rm min} < E < E_{\rm t}$ where we note $E_{\rm min} \equiv 2E_{\rm t}$. 
In this limited range of $E$, 
there are two real and positive solutions for equation (\ref{eq_r1r2}), or equivalently, 
\begin{equation} \label{eq_linear}
2GMr_{\rm t} {\left( \frac{r}{r_{\rm t}} \right)}^3 
- 2r_{\rm t}^2 \left( E - E_{\rm min} \right) {\left( \frac{r}{r_{\rm t}} \right)}^2 
+ L^2 = 0 , 
\end{equation}
if and only if 
\begin{equation}
0 < D < 2,
\end{equation}
where 
\begin{equation}
D = \frac{27 G^2 M^2 L^2}{4 r_{\rm t}^4 {\left( E -E_{\rm min} \right)}^3} . 
\end{equation}
In this allowed region, two real and positive solutions for 
equation (\ref{eq_linear}) are as follows:
\begin{equation}
r_i = \frac{2r_{\rm t}^2}{3GM} \left( E -E_{\rm min} \right) x_i  
\;\; ( i= {\rm apo}\; {\rm or} \; {\rm peri}; \;r_{\rm apo} > r_{\rm peri} )  ,
\end{equation}
with $x_{\rm apo}$ and $x_{\rm peri}$ given, respectively, by
\begin{equation}
x_{\rm apo} = \frac{1}{2} + \cos \theta , \hspace{2ex} {\rm and} \hspace{2ex}
x_{\rm peri} = \frac{1}{2} + \cos \left[ \frac{4\pi}{3} + \theta \right]  ,
\end{equation}
thus
\begin{equation}
e = \frac{\cos \theta - \cos \left[ \frac{4\pi}{3} + \theta \right]}
{1 + \cos \theta + \cos \left[ \frac{4\pi}{3} + \theta \right]} , 
\end{equation}
where  
\begin{equation}
    \theta = 
    \begin{cases}
        \frac{1}{3} \tan^{-1} \left( \frac{\sqrt{2D-D^2}}{1-D} \right) ,
        & \text{if $0 < D < 1$} \\
        \frac{1}{3} \left[ \tan {}^{-1} \left( \frac{\sqrt{2D-D^2}}{1-D} \right) + \pi \right] ,
        & \text{if $1 < D < 2$.} 
    \end{cases}
\end{equation}
Consequently, $D$ has a one-to-one correspondence to $e$, so with 
$\theta$, $x_{\rm apo}$, and $x_{\rm peri}$. Use of these quantities gives 
\begin{equation}
L^2 = \frac{4r_{\rm t}^4 D}{27 G^2 M^2} {\left( E -E_{\rm min} \right)}^3, 
\end{equation}
\begin{equation}
T_r = 2\sqrt{3} \frac{r_{\rm t}^2 \sqrt{E-E_{\rm min}}}{GM}  
\int_{x_{\rm peri}}^{x_{\rm apo}} \frac{x dx}{\sqrt{-x^3 + \frac{3}{2} x^2 - \frac{D}{4}}}, 
\end{equation}
and 
\begin{multline}
{\left(\frac{\partial L^2}{\partial e}\right)}_{E} = 
- \frac{4 r_{\rm t}^4 \sqrt{2D-D^2} }{9G^2M^2} (E -E_{\rm min})^3 \\
\times
\frac{{(1 + \cos \theta + \cos \left[ \frac{4\pi}{3} + \theta \right])}^2}
{\sin \theta \left( 1 + 2 \cos \left[ \frac{4\pi}{3} + \theta \right] \right) 
- \sin \left[ \frac{4\pi}{3} + \theta \right] \left( 1 + 2 \cos \theta \right)} . 
\end{multline}
By these expressions, we obtain 
\begin{multline}
n_{\beta}(E,e) = 24\sqrt{3} \pi^2 \frac{r_{\rm t}^2}{GM} 
{\left( \frac{4r_{\rm t}^4}{27 G^2 M^2} \right)}^{1-\beta} \\
\times D^{-\beta} \sqrt{2D-D^2} 
{\left( E -E_{\rm min} \right)}^{\frac{7}{2}-3\beta}  \\
\times 
\frac{{(1 + \cos \theta + \cos \left[ \frac{4\pi}{3} + \theta \right])}^2}
{\sin \theta \left( 1 + 2 \cos \left[ \frac{4\pi}{3} + \theta \right] \right) 
- \sin \left[ \frac{4\pi}{3} + \theta \right] \left( 1 + 2 \cos \theta \right)} \\
\times
\int_{x_{\rm peri}}^{x_{\rm apo}} \frac{x dx}{\sqrt{-x^3 + \frac{3}{2} x^2 - \frac{D}{4}}} .
\end{multline}
Since $\theta$, $D$, $x_{\rm apo}$, and $x_{\rm peri}$ depend only on $e$, 
$n_{\beta}(E,e)$ is separable in $E$ and $e$, so that  
\begin{multline}
N_{\beta}(e) = 24\sqrt{3} \pi^2 \frac{r_{\rm t}^2}{GM} 
{\left( \frac{4r_{\rm t}^4}{27 G^2 M^2} \right)}^{1-\beta} \\
\times D^{-\beta} \sqrt{2D-D^2} 
\left[ \int g(E) {\left( E -E_{\rm min} \right)}^{\frac{7}{2}-3\beta} dE \right] \\
\times 
\frac{{(1 + \cos \theta + \cos \left[ \frac{4\pi}{3} + \theta \right])}^2}
{\sin \theta \left( 1 + 2 \cos \left[ \frac{4\pi}{3} + \theta \right] \right) 
- \sin \left[ \frac{4\pi}{3} + \theta \right] \left( 1 + 2 \cos \theta \right)} 
\\
\times
\int_{x_{\rm peri}}^{x_{\rm apo}} \frac{x dx}{\sqrt{-x^3 + \frac{3}{2} x^2 - \frac{D}{4}}} . 
\end{multline}
Therefore, the shape of $N_{\beta}(e)$ is not affected by $E$ or $g(E)$, 
like the point mass model and the truncated model with any $\gamma$. 
The results of $N_{\beta}(e)$ in the linear potential model are shown on the left 
panel of Figure \ref{Fig linear+SIS-beta}.  We see that $N_{\beta}(e)$ 
is a monotonically increasing $e$-distribution for $0.52 < \beta < 1$, 
and $N_{\beta}(e)$ has a hump-like $e$-distribution with a single peak 
for $\beta < 0.5$. In particular, $N_{\beta=0}(e)$ reaches its maximum 
at $e_{\rm peak}=0.43$.  In the intermediate range of $0.5 < \beta < 0.52$, 
$N_{\beta}(e)$ shows something like a trapezoidal $e$-distribution, 
which shows a monotonically increasing $e$-distribution for  $0 < e < e_{\rm peak}$ 
and a more or less flat behavior for  $e_{\rm peak} < e < 1$, 
where $e_{\rm peak} \simeq 0.8$. 

\subsection{Singular isothermal model ($\gamma = 2$)} \label{SIS}
One of the most strong constraints on the gravitational potential of the halo 
is that it has to be consistent with the observed flat rotation curve of 
galaxy disk. In this sense, the truncated singular isothermal model, 
which automatically reproduces the flat rotation curve in the radial range 
of $0<r<r_{\rm t}$, is said to be one of the simple and realistic models. 
The density profile of this model is given by 
\begin{equation}
    \rho(r) = 
    \begin{cases}
        \frac{M}{4\pi r_{\rm t}^3} {\left( \frac{r}{r_{\rm t}} \right)}^{-2}, & \text{if $r < r_{\rm t}$} \\
        0,  & \text{otherwise,} 
    \end{cases}
\end{equation}
which corresponds to $\gamma = 2$ in equation (\ref{power-law}). 
The gravitational potential arising from this density profile is given by 
\begin{equation}
     V(r) = 
     \begin{cases}
     - \frac{GM}{r_{\rm t}} + \frac{GM}{r_{\rm t}} \ln \left( \frac{r}{r_{\rm t}} \right), & \text{if $r < r_{\rm t}$} \\ 
     -\frac{GM}{r}, &\text{otherwise}. 
     \end{cases} 
\end{equation}
We consider only stars with $E<E_{\rm t}\equiv-GM/r_{\rm t}$, 
which guarantees the stars to be confined inside the truncated radius $r_{\rm t}$.  
Thus, bound orbits within the truncated sphere are allowed if $-\infty < E < E_{\rm t}$. 
In this range of $E$, there are two real and positive solutions for equation (\ref{eq_r1r2}), 
or equivalently, 
\begin{equation} 
-2(E r_{\rm t}^2 + GMr_{\rm t}) {\left( \frac{r}{r_{\rm t}} \right)}^2 
+ 2GMr_{\rm t} {\left( \frac{r}{r_{\rm t}} \right)}^2 \ln \left( \frac{r}{r_{\rm t}} \right) + L^2 = 0 ,
\end{equation}
if and only if 
\begin{equation}
0 < D < \frac{1}{2 \exp (1)} ,  
\end{equation}
where 
\begin{equation}
D \equiv \frac{L^2}{2GMr_{\rm t} \exp \left[ 2 \left( 1 + \frac{r_{\rm t} E}{GM} \right) \right]} . 
\end{equation}
In this allowed region, two real and positive solutions are as follows:
\begin{equation}
r_i = r_{\rm t} \exp \left( 1 + \frac{r_{\rm t} E}{GM} \right) x_i  ,
\;\; (i= {\rm apo}\; {\rm or} \; {\rm peri}; \; r_{\rm apo} > r_{\rm peri} )
\end{equation}
with $x_{\rm apo}$ and $x_{\rm peri}$ are the solutions for 
\begin{equation} \label{eqSIS}
x^2 \ln x + D = 0 .  
\end{equation}
By this equation, $D$ has a one-to-one correspondence to $e$, so with 
$x_{\rm peri}$ and $x_{\rm apo}$. Use of these quantities gives 
\begin{equation}
 T_r = \sqrt{\frac{2 r_{\rm t}^3}{GM}} \exp \left( 1 + \frac{r_{\rm t} E}{GM} \right)
\int_{x_{\rm peri}}^{x_{\rm apo}} \frac{x dx}{\sqrt{-D - x^2 \ln x}} , 
\end{equation}
\begin{equation}
L^2 = 2 GMr_{\rm t} D \exp \left[ 2 \left( 1 + \frac{r_{\rm t} E}{GM} \right) \right], 
\end{equation}
and 
\begin{multline}
{\left( \frac{\partial L^2}{\partial e} \right)}_E = 
- GMr_{\rm t} \exp \left[ 2 \left( 1 + \frac{r_{\rm t} E}{GM} \right) \right] 
\\
\times
\frac{ {(x_{\rm apo} + x_{\rm peri})}^2 }{x_{\rm apo} x_{\rm peri}} 
{\left[ \frac{1}{x_{\rm apo}^2  - 2D} 
       - \frac{1}{x_{\rm peri}^2 - 2D} 
\right]}^{-1} .
\end{multline}
Consequently, we obtain 
\begin{multline} 
n_{\beta}(E,e) = 
4 \sqrt{2} \pi^2 \sqrt{ GMr_{\rm t}^5 } 
{\left[ 2GMr_{\rm t} D \right]}^{-\beta} 
\\
\times
\exp \left[ (3-2\beta) \left( 1 + \frac{r_{\rm t} E}{GM} \right) \right] \\
\times
\frac{ {(x_{\rm apo} + x_{\rm peri})}^2 }{x_{\rm apo} x_{\rm peri}} 
{\left[ \frac{1}{x_{\rm apo}^2  - 2D} 
       - \frac{1}{x_{\rm peri}^2 - 2D} 
\right]}^{-1} 
\\
\times
\int_{x_{\rm peri}}^{x_{\rm apo}} \frac{x dx}{\sqrt{-D - x^2 \ln x}} . 
\end{multline}
Since $D$, $x_{\rm peri}$, and $x_{\rm apo}$ depend only on $e$, 
$n_{\beta}(E,e)$ is separable in $E$ and $e$, so that 
\begin{multline} 
N_{\beta}(e) = 
4 \sqrt{2} \pi^2 \sqrt{ GMr_{\rm t}^5 } 
{\left[ 2GMr_{\rm t} D \right]}^{-\beta} 
\\
\times
\left[ \int 
g(E) \exp \left[ (3-2\beta) \left( 1 + \frac{r_{\rm t} E}{GM} \right) \right] dE
\right]
\\
\times
\frac{ {(x_{\rm apo} + x_{\rm peri})}^2 }{x_{\rm apo} x_{\rm peri}} 
{\left[ \frac{1}{x_{\rm apo}^2  - 2D} 
       - \frac{1}{x_{\rm peri}^2 - 2D} 
\right]}^{-1} 
\\
\times
\int_{x_{\rm peri}}^{x_{\rm apo}} \frac{x dx}{\sqrt{-D - x^2 \ln x}} . 
\end{multline}
Thus, the shape of $N_{\beta}(e)$ is not affected by $g(E)$, 
like the point mass model and the truncated model with any $\gamma$.
The results of $N_{\beta}(e)$ in the singular isothermal model are shown on  
the right panel of Figure \ref{Fig linear+SIS-beta}.  We see that $N_{\beta}(e)$ 
shows a monotonically increasing $e$-distribution for $\beta > 0.45$, while 
having a single peak for $\beta < 0.45$.
 \label{lastpage}

\end{document}